\documentclass[12pt,a4paper]{article}
\usepackage[latin1]{inputenc}
\usepackage{a4wide}
\usepackage{amsfonts}
\usepackage{amssymb}
\usepackage{amsmath}
\usepackage{ifpdf}
\ifpdf
\usepackage[pdftex]{graphicx}
\usepackage[pdftex,unicode,implicit]{hyperref}
\hypersetup{%
  pdftitle    = {Domain walls and instantons in N=1, d=4 supergravity},
  pdfkeywords = {Supersymmetry, supergravity, symmetry, gauge, branes },
  pdfauthor   = {Mechthild Hübscher, Patrick Meessen, Tomás Ortín},
  pdfcreator  = {pdf\LaTeXe\ with package \flqq hyperref\frqq},
  pdfproducer = {pdf\LaTeXe\ with package \flqq hyperref\frqq},
  pdfpagemode = None,  
  pdffitwindow= true,  
  unicode     = true,
  plainpages  = true,
  colorlinks  = true,  
  citecolor   = blue,
  urlcolor    = red,
  linkcolor   = black
}
\newcommand{\hepth}[1]{arXiv:{\tt
\href{http://www.arXiv.org/abs/hep-th/#1}{hep-th/#1}}}

\newcommand{\arxiv}[1]{{\tt
\href{http://www.arXiv.org/abs/#1}{arXiv:#1}}}
\else
  \usepackage[dvips]{graphicx}
  \usepackage{picinpar}
  \usepackage[unicode,implicit]{hyperref}
  \newcommand{\hepth}[1]{arXiv:{\tt hep-th/#1}}

  \newcommand{\arxiv}[1]{{\tt arXiv:#1}}
\fi
\makeatletter
\@addtoreset{equation}{section}
\makeatother

\makeindex
\begin{document}

\begin{flushright}
\small
IFT-UAM/CSIC-09-44\\
December  $13^{\rm th}$, 2009\\
\normalsize
\end{flushright}

\begin{center}

\vspace{.7cm}

{\LARGE {\bf Domain Walls and Instantons\\[.8cm] 
in\\[.8cm] 
$N=1$, $d=4$ Supergravity}}

\vspace{1.5cm}

\begin{center}

{\bf  Mechthild H\"ubscher, Patrick Meessen and Tom\'as Ort\'{\i}n} 
%
%
\vspace{1cm}

\textit{Instituto de F\'{\i}sica Te\'orica UAM/CSIC
Facultad de Ciencias C-XVI, \\
C.U. Cantoblanco, E-28049-Madrid, Spain\vskip 5pt}

\vspace{.5cm}

{e-mail: {\tt Mechthild.Huebscher , Patrick.Meessen , Tomas.Ortin @ uam.es}}

\end{center}

\vspace{2cm}

{\bf Abstract}

\begin{quotation}

  {\small 
    We study the supersymmetric sources of (multi-) domain-wall and (multi-)
    instanton solutions of generic $N=1,d=4$ supergravities, that is: the
    worldvolume effective actions for these supersymmetric topological
    defects.

    The domain-wall solutions naturally couple to the two 3-forms recently
    found as part of the $N=1,d=4$ tensor hierarchy (i.e.~they have two
    charges in general) and their tension is the absolute value of the
    superpotential section $\mathcal{L}$. The introduction of sources (we
    study sources with finite and vanishing thickness) is equivalent to the
    introduction of local coupling constants and results in dramatic changes
    of the solutions. Our results call for a democratic reformulation of
    $N=1,d=4$ supergravity in which coupling constants are, off-shell, scalar
    fields.

    The effective actions for the instantons are always proportional to a null
    coordinate (in the Wick-rotated scalar manifold) which is constant over
    the whole space in the instanton solution. We show their supersymmetry and
    find the associated supersymmetric (multi-) instanton solutions.

}

\end{quotation}

\end{center}

\newpage
\pagestyle{plain}

\tableofcontents

\newpage

\section{Introduction}

Over the last decades, the effective actions of supersymmetric extended
objects (\textit{branes}) have played a crucial role in many
developments. First of all, as source terms for the supergravity solutions
that describe the branes, they confirm the relation between $(p+1)$-form
potentials and $p$-branes. Often, the requirements of $\kappa$-symmetry and
gauge invariance of the worldvolume effective action of a given $p$-brane plus
the relation via duality with the effective actions of other branes lead to
the addition of worldvolume fields different from the embedding coordinates
which can be associated to the dynamics of the boundaries of other branes
ending on the $p$-brane's worldvolume. They also lead to the presence of the
potentials associated to other branes in the Wess-Zumino term. These
non-trivial elements of the action give a great deal of information about
possible intersections of branes and have led, for instance to the discoveries
of the Myers and enhan\c{c}on effects.

One of the fundamental ingredients of the $p$-branes worldvolume effective
actions is the knowledge of the $(p+1)$-form potentials of the supergravity
theory (their own existence and gauge and supersymmetry
transformations). While the lower-rank potentials are present from the onset
in the standard formulations of the supergravity theories, the higher-rank
potentials have to be found. In some cases this can be done using via
Hodge-duality, but this procedure becomes very complicated for $p=d-2$ (forms
coupling to strings, which are dual to scalars) and $p=d-1$ (forms coupling to
domain walls, which are dual to coupling constants, not all of which may be
known) and impossible for $p=d$ (forms coupling to spacetime-filling
branes). Therefore, the \textit{democratic} formulation of the $d=10$ type~II
supergravity theories \cite{Bergshoeff:2001pv} is necessarily incomplete,
although self-consistent and sufficient if one is mainly interested in
D-branes.  This has motivated the systematic seach for all the higher-rank
potentials of (in particular) the $d=10$ type~II supergravity theories
\cite{Bergshoeff:2005ac,Bergshoeff:2006qw} which can later be used for the
construction of the worldvolume effective theories of the correponding branes
\cite{Bergshoeff:2006ic}.

It is, therefore, very interesting and useful to find the effective actions of
as many supersymmetric extended objects as possible. Since much less is known
about the effective actions of the $p$-branes of lower-dimensional
supergravities, in this article we will focus on the supersymmetric objects of
$N=1$ $d=4$ supergravity, which correspond to the supersymmetric solutions
recently classified and characterized in
Refs.~\cite{Gran:2008vx,Ortin:2008wj}, plus the supersymmetric instantons
which, being solutions of the Wick-rotated theory, were not studied in those
references. In particular, we will focus on the effective actions of
supersymmetric domain walls and instantons, since in $N=1,d=4$ supergravity
there are no supersymmetric black holes (0-branes) nor spacetime-filling
branes (3-branes) and the supersymmetric string solutions are essentially
identical to those of the $N=2$ case, treated in \cite{Bergshoeff:2007ij} and
their worldsheet effective actions should not be too different; in particular,
their tensions should be given by the momentum maps associated to the
symmetries involved.

Domain-wall solutions in $N=1$ $d=4$ sugra were first found in
Ref.~\cite{Cvetic:1992bf}, and extensively discussed in
Ref.~\cite{Cvetic:1996vr}, but the 3-forms to which they must couple are
lacking from the ordinary formulation of $N=1,d=4$ supergravity.  Recently,
the consistent addition of 2-, 3- and 4-form potentials to $N=1,d=4$
supergravity was systematically investigated in Ref.~\cite{Hartong:2009az}
using the general form of the 4-dimensional tensor hierarchy
\cite{Cordaro:1998tx,deWit:2005ub,deWit:2008ta} determined in
Ref.~\cite{Bergshoeff:2009ph} and supersymmetrizing it. 

Two types of 3-form potentials were found: the 3-form potentials associated to
the possible gaugings of isometries of the scalar manifold (and, therefore,
associated or ``dual''\footnote{To be precise, 3-forms are not dual to the
  deformation parameters (coupling constants, mass parameters etc)
  themselves. As shown in Ref.~\cite{Bergshoeff:2009ph} the Hodge dual of each
  3-form field strength is the derivative of the scalar potential with respect
  to the associated deformation parameter. Only when the potential is just a
  cosmological constant (the square of a deformation parameter), 3-forms are
  dual in a strict sense to deformation parameters.} to the corresponding
components of the embedding tensor) on the one hand, and two 3-forms not
related to such gaugings. The interesting difference between these two sets of
3-forms is that the supersymmetry variation of the latter contain the
gravitino, whereas the former do not.  This difference is enough to discard
the first set as possible 3-forms to couple electrically to dynamical domain
walls as the kinetic term of the domain-wall effective action should contain a
standard Nambu-Goto term, whose supersymmetry variation always contains the
gravitino. The standard Bose-Fermi matching arguments \cite{Achucarro:1987nc}
predict that a $\kappa$-symmetric 3-brane action can be constructed without
additional worldvolume fields using the second set of 3-forms.

These last two 3-form potentials, not predicted by the bosonic tensor hierarchy,
denoted by $C$ and $C^{\prime}$ in Ref.~\cite{Hartong:2009az} are
distinguished by the fact that $C$, must have a vanishing field strength
($dC=0$, i.e.~it is dual to nothing), whereas the field strength of
$C^{\prime}$ is dual to the part of the scalar potential of $N=1,d=4$
supergravity that depends on the superpotential (and not on the gaugings).
These conditions, required by the closure of the local supersymmetry algebra
on the 3-forms, were interpreted in Ref.~\cite{Hartong:2009az} as follows:
$C^{\prime}$ must by associated to a deformation parameter $g$ that can be
made manifest by rescaling the superpotential with $g$. This coupling constant
$g$, then, is associated to the presence of the superpotential in the theory:
when $g=0$ there is no superpotential. As for $C$, it was conjectured that
$N=1,d=4$ supergravity may admit another, yet unknown, deformation to which
$C$ would be associated.

As we are going to show in this paper, both 3-forms play indeed a very similar
role: if combine them into a complex 3-form $\mathcal{C}=C^{(1)}+iC^{(2)}$,
then $\mathcal{C}$ is associated to a complex deformation parameter
$g^{(1)}+ig^{(2)}$ that can be made manifest by rescaling with it the
superportential. The field strength $d\mathcal{C}$ has to be the dual of the
the part of the scalar potential of $N=1,d=4$ supergravity that depends on the
superpotential, but multiplied by $g^{(1)}+ig^{(2)}$, so its real or imaginary
part can be made dual to nothing if the real or imaginary parts of the complex
coupling constant are set to zero. Then, the reason why $C$ and $C^{\prime}$
in Ref.~\cite{Hartong:2009az} seemed to play a very different role was due to
the fact that the standard coupling of the superpotential to the rest of the
supergravity theory is not the most general one. The most general one is
obtained by multiplying everywhere the superpotential by a complex phase. This
generalization does not modify the bosonic part of the theory, but it is
noticeable in the couplings to fermions. This freedom is reflected in the
existence of another deformation parameter which justifies the existence of a
second 3-form (the old $C$). In practice, this freedom may not have important
physical effects because it can always be absorbed in redefinitions of the
superpotential but confirms the standard lore that there must be a deformation
parameter for every $(d-1)$-form potential.

The construction of an effective action for a domain-wall charged with respect
to both 3-forms is sraightforward. As mentioned before, no additional
worldvolume fields are needed for Bose-Fermi matching which implies, in
particular, that there is no Born-Infeld vector associated to strings ending
on the domain wall. This agrees with the absence of supersymmetric solutions
describing them \cite{Gran:2008vx,Ortin:2008wj}: in the only supersymmetric
solutions describing strings and domain walls, these are parallel
\cite{Gutowski:2001ap}. The worldvolume 2-forms that may describe them are
non-dynamical\footnote{In $N=2,d=4$ supergravity the situation is completely
  different: a worldvolume vector field is necessary for Bose-fermi matching
  \cite{Duff:1992hu}, so strings can end on $N=2,d=4$ domain walls. Since
  there are several kinds of $N=2,d=4$ strings and probably also of domain
  walls (this has not been studied yet because the 3-forms of $N=2,d=4$
  supergravity and their supergravity transformations are not known) the
  construction of the effective actions is not trivial. The spacetime-filling
  branes of $N=2,d=4$ supergravity may also have interesting interactions with
  other $N=2,d=4$ branes for similar reasons.}.

A problem arrises when we couple said domain-wall effective action to the
(bulk) $N=1,d=4$ supergravity action to use it as a source, as we lack a fully
democratic formulation of the action including the 3-forms. However, on
general grounds \cite{Bergshoeff:2000zn} the coupling of 3-forms to the rest
of the supergravity action can always be constructed as follows: promoting the
coupling constant $g$ to a coupling \textit{function} $g(x)$ and adding a
Lagrange-multiplier term enforcing the constraint $dg=0$. This
Lagrange-multiplier term is of the form $C\wedge dg$ where $C$ is the
$(d-1)$-form associated to the deformation parameter $g$. The promotion of $g$
to a function $g(x)$ breaks gauge and supersymmetry invariance by terms of the
form $\Delta\wedge dg$, but this can be compensated by assigining to $C$ the
transformation rules $\delta C= -\Delta$. in aour case, the coupling constant
was introduced in the action by rescaling with it the superpotential and,
thus, appears multiplying the standard scalar potential $V_{\rm
  new}=g^{2}(x)V_{\rm old}$. Then, the coupling constant modulates the scalar
potential which has profound implications for the solutions.

As usual with supersymmetric configurations, there are first-order equations
(Bogomol'nyi equations, flow equations etc.) which imply some or all the
(sourceless) equations of motion so, in order to find supersymetric solutions,
it is enough to solve these first-order equations and then just a few (or no)
equations of motion. This is guaranteed by the Killing spinor identities
\cite{Kallosh:1993wx,Bellorin:2005hy} or, alternatively, by the integrability
conditions of the Killing spinor equations. A remarkable fact of our
construction is that the modified (with a spactime-dependent coupling $g(x)$)
first-order equations now imply the same equations of motion \textit{with
  sources}. This fact supports the consistency of our model and calls for the
construction of a fully supersymmetric and democratic action for $N=1,d=4$
supregravity.

The instantons we are after are modelled on the type IIB D-instanton
\cite{Gibbons:1995vg}:\footnote{ It would be more consistent to talk about
  D-instanton-like solutions or $\sigma$-model instantons, but we shall refer
  to them plainly as instantons as no confusion should arrise.} they are
solutions to Wick rotated $N=1$ $d=4$ sugra with a flat metric.  

As is discussed in Ref.~\cite{Gibbons:1995vg}, and more recently in
Ref.~\cite{Mohaupt:2009iq}, the effect of the Wick rotation on the scalar
manifold is that it is no longer K\"ahler, but rather para-K\"ahler, which is
a real manifold of split signature; such instanton solutions can however be
discussed for any theory of gravity coupled to a non-linear $\sigma$-model as
long as the metric of the $\sigma$-model is pseudo-Riemannian.  As is
well-known, these $\sigma$-model instantons correspond to null-geodesics on
the scalar manifold
\cite{Neugebauer:1969wr,Breitenlohner:1987dg,Clement:1986bt}.  An effective
action, or a source, for these instantons can be found by observing that the
effective action should just be some function of the scalars evaluated at the
location of the instanton. This leads to the conclusion that the source term
is the coordinate orthonormal to the twistfree congruence of the
null-geodesic.

This discussion is applicable to all instantons, but we are interested in
supersymmetric instantons and we will show that this kind of effective action
is also invariant under properly Wick-rotated supersymmetry transformation
rules.

The outline of this paper is as follows: in Section~\ref{sec:N1d4} we will
give a brief outline of $N=1,d=4$ supergravity coupled to chiral
supermultiplets.  In Section~\ref{sec:DomWall} we will treat the domain walls,
first from the bulk perspective in Section~\ref{sec-SSDWS}.
Section~\ref{sec:WorldVol} will be dedicated to the construction of the
worldvolume action which will be used as the source term;
Section~\ref{sec:Shebang} will then be dedicated to the discussion of the
change brought about by the introduction of the coupling function, needed for
the consistency of the construct. In Section~\ref{sec:Example}, then, these
changes will be illustrated by means of a simple example.

Section~\ref{sec:instantons} is dedicated to the instantons and starts off by
a general discussion of instantons and source-terms in
Section~\ref{sec:GenInst}. Supersymmetric instantons and their sources are
treated in Section~\ref{sec:SusyInst}, followed by some examples in
Sections~\ref{sec-instantonexamples1} and
\ref{sec-instantonexamples2}. Section~\ref{sec:conclusions} contains our
conclusions and outlook for future work.

\section{$N=1,d=4$ Supergravity coupled to $n_{C}$ chiral multiplets}
\label{sec:N1d4}

The theory we are going to work with consists of the supergravity multiplet
with one graviton $e^{a}{}_{\mu}$ and one chiral gravitino $\psi_{\mu}$ and
$n_{C}$ chiral multiplets with as many chiral dilatini $\chi^{i}$ and complex
scalars $Z^{i}$, $i=1,\cdots n_{C}$ parametrizing a K\"ahler-Hodge manifold
with K\"ahler potential $\mathcal{K}(Z,Z^{*})$. The couplings are dictated by
$\mathcal{K}$ and by the holomorphic superpotential $W(Z)$, which appears in
the theory through the covariantly holomorphic\footnote{ The K\"ahler
  connection 1-form $Q$ is defined by
\begin{equation}
\label{eq:K1form}
\mathcal{Q} \equiv {\textstyle\frac{1}{2i}}dZ^{i}\partial_{i}\mathcal{K} 
+\mathrm{c.c.}
\end{equation}
and, therefore, we have
  \begin{equation}
\mathcal{D}_{i^{*}}\mathcal{L} 
= 
(\partial_{i^{*}}+i\mathcal{Q}_{i^{*}})\mathcal{L}
=
e^{\mathcal{K}/2}\partial_{i^{*}}( e^{-\mathcal{K}/2}\mathcal{L})
=
e^{\mathcal{K}/2}\partial_{i^{*}}W
=
0\, .
\end{equation}
} section $\mathcal{L}(Z,Z^{*})$
of K\"ahler weight $(1,-1)$ defined by

\begin{equation}
\label{eq:N1defL}
  \mathcal{L}(Z,Z^{*})\equiv W(Z) e^{\mathcal{K}/2}\, .
\end{equation}

The action for the bosonic fields is

\begin{equation}
\label{eq:actionN1ungauged}
 S  =  {\displaystyle\int} d^{4}x \sqrt{|g|}
\left[R +2\mathcal{G}_{ij^{*}}\partial_{\mu}Z^{i}
\partial^{\mu}Z^{*\, j^{*}} 
-V(Z,Z^{*})
\right]\, ,
\end{equation}

\noindent
where the scalar potential $V_{\rm u}(Z,Z^{*})$ is entirely determined by the
superpotential $\mathcal{L}$ through the expression\footnote{It is customary
  to introduce a coupling constant, $g$, into the potential, or in the
  superpotential $W$, but as it can be reinstated trivially and not really
  needed in this section, we will obviate it for the moment; it will, however,
  be reinstated in csection~\ref{sec:Shebang}.}

\begin{equation}
  V(Z,Z^{*}) = 
  -24 |\mathcal{L}|^{2} 
  +8\mathcal{G}^{ij^{*}}\mathcal{D}_{i}\mathcal{L} 
  \mathcal{D}_{j^{*}}\mathcal{L}^{*}\, , 
\end{equation}

\noindent
where

\begin{equation}
\mathcal{D}_{i}\mathcal{L} 
= 
(\partial_{i}+i\mathcal{Q}_{i})\mathcal{L}
=
e^{\mathcal{-K}/2}\partial_{i}( e^{\mathcal{K}/2}\mathcal{L})
=
e^{-\mathcal{K}/2}\partial_{i}(e^{\mathcal{K}}W)\, .
\end{equation}

\noindent
We will also use the ``fermion shift''

\begin{equation}
\label{eq:fermionshift}
\mathcal{N}^{i} \equiv 2 \mathcal{G}^{ij^{*}}\mathcal{D}_{j^{*}}\mathcal{L}^{*}\, ,
\end{equation}

\noindent
that appear in the chiralino supersymmetry transformations, in terms of which
the scalar potential takes the form

\begin{equation}
V(Z,Z^{*}) = 
-24 |\mathcal{L}|^{2} +2\mathcal{G}_{ij^{*}}\mathcal{N}^{i}\mathcal{N}^{*\, j^{*}}
\, . 
\end{equation}

For vanishing fermions, the \textit{standard} fermionic supersymmetry
transformations take the form

\begin{eqnarray}
\delta_{\epsilon}\psi_{\mu} & = & 
\mathcal{D}_{\mu}\epsilon +i\mathcal{L}\gamma_{\mu}\epsilon^{*}
=
\left[\nabla_{\mu} 
+{\textstyle\frac{i}{2}}\mathcal{Q}_{\mu}\right]\epsilon 
+i\mathcal{L}\gamma_{\mu}\epsilon^{*}\, ,
\label{eq:gravisusyruleN1}\\
& & \nonumber \\
\delta_{\epsilon}\chi^{i} & = & 
i\not\!\partial Z^{i}\epsilon^{*}+\mathcal{N}^{i}\epsilon\, ,
\label{eq:dilasusyruleN1}
\end{eqnarray}

\noindent
where $\mathcal{Q}_{\mu}$ is the pullback of the K\"ahler connection 1-form 

\begin{equation}
\mathcal{Q}_{\mu} = \partial_{\mu}Z^{i}\mathcal{Q}_{i} +\mathrm{c.c.}  
\end{equation}

Observe that replacing $\mathcal{L}$ by, for instance,
$\tfrac{1}{\sqrt{2}}(1+i)\mathcal{L}$, leaves the bosonic action invariant but
modifies the fermion shifts in the fermion supresymmetry transformations. This
change can be seen as a field redefinition since the phase
$\tfrac{1}{\sqrt{2}}(1+i)$ can always be reabsorbed into $\mathcal{L}$,
bringing the supersymmetry transformations back to the \textit{standard}
form. This will not be possible after the introduction of a local coupling
constant in the coming sections and, therefore, it is important to notice this
possibility.

The supersymmetry transformation rules for the bosonic fields for
vanishing fermions are

\begin{eqnarray}
\label{eq:susytranseN1}
\delta_{\epsilon} e^{a}{}_{\mu} & = &  
-{\textstyle\frac{i}{4}} \bar{\psi}_{\mu}\gamma^{a}\epsilon^{*}
+\mathrm{c.c.}\, ,\\
& & \nonumber \\
\delta_{\epsilon} Z^{i} & = & 
{\textstyle\frac{1}{4}} \bar{\chi}^{i}\epsilon\, .
\label{eq:susytransZN1}
\end{eqnarray}

We denote (the l.h.s.~of) the bosonic equations of motion by

\begin{equation}
\mathcal{E}_{a}{}^{\mu}\equiv 
-\frac{1}{2\sqrt{|g|}}\frac{\delta S}{\delta e^{a}{}_{\mu}}\, ,
\hspace{.5cm}
\mathcal{E}^{i} \equiv -\frac{\mathcal{G}^{ij^{*}}}{2\sqrt{|g|}}
\frac{\delta S}{\delta Z^{*j^{*}}}\, ,
\end{equation}

\noindent
and they take the form

\begin{eqnarray}
\mathcal{E}_{\mu\nu} & = & 
G_{\mu\nu}
+2\mathcal{G}_{ij^{*}}[\partial_{\mu}Z^{i} \partial_{\nu}Z^{*\, j^{*}}
-{\textstyle\frac{1}{2}}g_{\mu\nu}
\partial_{\rho}Z^{i}\partial^{\rho}Z^{*\, j^{*}}]
+{\textstyle\frac{1}{2}}g_{\mu\nu}V\, ,
\label{eq:Emn}\\
& & \nonumber \\
\mathcal{E}_{i} 
& = & 
\mathcal{G}_{ij^{*}}\nabla^{2} Z^{*\, j^{*}}
+{\textstyle\frac{1}{2}}\partial_{i}V\, .
\label{eq:Ei}
\end{eqnarray}

A compact expression for the derivative of the potential is

\begin{equation}
{\textstyle\frac{1}{2}}\mathcal{G}^{ij^{*}}\partial_{j^{*}}V =   
-4\mathcal{L}\mathcal{N}^{i} +\mathcal{N}^{*\,
  j^{*}}\mathcal{D}_{j^{*}}\mathcal{N}^{i}\, .
\end{equation}

\section{Supersymmetric domain walls}
\label{sec:DomWall}

\subsection{Sourceless supersymmetric domain-wall solutions}
\label{sec-SSDWS}

In this section we are going to review the standard supersymmetric domain-wall
solutions of $N=1,d=4$ supergravity that can be found in the literature. They
solve the equations of motion derived from the supergravity action alone,
without any additional sources and can be thought of as describing the
gravitational and scalar fields far from where the possible sources are
placed.

The metric of a 4-dimensional domain-wall solution can always be brought into
the form

\begin{equation}
\label{eq:domainwallmetric}
ds^{2} = H\eta_{\mu\nu}dx^{\mu}dx^{\nu} = H[\eta_{mn}dx^{m}dx^{n} -dy^{2}]\, ,
\hspace{1cm}
m,n=0,1,2\, ,  
\end{equation}

\noindent
where $H$ is a function of the transverse coordinate $x^{3}\equiv y$ only.

In the Vierbein basis 

\begin{equation}
e^{a}{}_{\mu} = H^{1/2}\delta^{a}{}_{\mu}\, ,
\hspace{1cm}  
e_{a}{}^{\mu} = H^{-1/2}\delta_{a}{}^{\mu}\, ,
\end{equation}

\noindent
the  components of the spin connection are 

\begin{equation}
  \omega_{\mu}{}^{bc} =
\eta^{3[b}e^{c]}{}_{\mu} \partial_{\underline{y}}\log{H}\, .  
\end{equation}

\noindent
Using these in the world-volume components of the gravitino supersymmetry
transformation rule ($\mu=m=0,1,2$) Eq.~(\ref{eq:gravisusyruleN1}) and
assuming that both the complex scalars $Z^{i}$ and the Killing spinors
$\epsilon$ only depend on $y$, we find the unbroken supersymmetry condition

\begin{equation}
  i\gamma^{3} \partial_{\underline{y}}H^{-1/2}\epsilon + 2\mathcal{L}\epsilon^{*} =0\, ,    
\end{equation}

\noindent
which can only be consistently imposed if the metric function $H$ satisfies
the flow equation

\begin{equation}
\label{eq:sourcelessflowH}
\partial_{\underline{y}}H^{-1/2} = \pm 2 |\mathcal{L}|\, .  
\end{equation}

\noindent
When this equation is satisfied, the unbroken supersymmetry condition becomes
the 1/2-supersymmetry-preserving projector\footnote{The Killing spinor
  equation associated to the transverse direction has a complicated form, but
  can always be solved without requiring any further conditions.}

\begin{equation}
i\gamma^{3} (e^{-i\alpha/2}\epsilon) \pm (e^{-i\alpha/2}\epsilon)^{*} =0\, ,    
\end{equation}

\noindent
where we have defined the phase

\begin{equation}
e^{i\alpha} \equiv \mathcal{L}/|\mathcal{L}|\, .  
\end{equation}

Multiplying the above projector by $-\gamma^{012}$ and using the chirality of
the spinors $-i\gamma^{0123}\epsilon=\gamma_{5}\epsilon=-\epsilon$ we can
rewrite in the characteristic form of a domain-wall supersymmetry projector:

\begin{equation}
\label{eq:dwsusyprojector}
(e^{-i\alpha/2}\epsilon) \pm i\gamma^{012}(e^{-i\alpha/2}\epsilon)^{*} =0\, .    
\end{equation}

Using now the above projector into the chiralino supersymmetry transformation
rule Eq.~(\ref{eq:dilasusyruleN1}) we find a second flow equation for the
complex scalars \cite{Cvetic:1991vp}:

\begin{equation}
\label{eq:sourcelessflowZi}
\partial_{\underline{y}}Z^{i}= \pm e^{i\alpha}\mathcal{N}^{i}H^{1/2}\, .
\end{equation}
 
It is enough to impose the first-order flow equations
(\ref{eq:sourcelessflowH}) and (\ref{eq:sourcelessflowZi}) on $H,Z^{i}$ to
have a solution of the second-order supergravity equations of motion, as can
be dediced from the corresponding Killing spinor identities
\cite{Kallosh:1993wx,Bellorin:2005hy}. In particular, the worldvolume
components of the Einstein equations\footnote{The non-vanishing components of
  the Einstein tensor of the domain-wall metric
  Eq.~(\ref{eq:domainwallmetric}) are
 \begin{equation}
    \begin{array}{rcl}
      G_{\underline{m}\underline{n}}
& = & 
\eta_{mn}[H^{-1}\partial^{2}_{\underline{y}}H -\tfrac{3}{4}H^{-2}
(\partial_{\underline{y}}H)^{2}]\, ,
\\
& & \\
G_{\underline{y}\underline{y}}
& = & 
-\tfrac{3}{4}H^{-2}
(\partial_{\underline{y}}H)^{2}\, .\\
    \end{array}
  \end{equation}
} (\ref{eq:Emn}) can be written in the form

\begin{equation}
  \mathcal{E}_{\underline{m}\underline{n}} \sim  \eta_{mn} 
\partial_{\underline{y}} [\partial_{\underline{y}}H^{-1/2} \mp 2
|\mathcal{L}|]=0\, ,
\end{equation}

\noindent
the transverse components in the form

\begin{equation}
  \mathcal{E}_{\underline{y}\underline{y}} \sim  
-3 H[\partial_{\underline{y}}H^{-1/2} \mp 2
|\mathcal{L}|]^{2} 
+\mathcal{G}_{ij^{*}}[\partial_{\underline{y}}Z^{i}
\mp e^{i\alpha}\mathcal{N}^{i}H^{1/2}]
[\partial_{\underline{y}}Z^{*j^{*}}
\mp e^{-i\alpha}\mathcal{N}^{*\, i^{*}}H^{1/2}]
=0\, ,
\end{equation}

\noindent
and the equations of motion of the scalars (\ref{eq:Ei}) can be written in a
similar, more complicated form, proportional to the flow equations as well.

\subsection{Supersymmetric sources:  world-volume effective actions}
\label{sec:WorldVol}

Charged 4-dimensional domain walls must couple to 3-forms. If their effective
action (our candidate for a source term for domain-wall solutions) is going to
be $\kappa$-symmetric it must be invariant under supersymmetry transformations
and this requires that it consists of a kinetic Nambu-Goto-like term and a
Wess-Zumino-like term containing the 3-form. As discussed in the introduction,
no worldvolume fields should be needed in addition to the embedding
coordinates and fermions. Furthermore, the supersymmetry transformations of
the 3-form must contain the gravitino in order to have a chance to cancel the
supersymmetry transformations of the metric in the kinetic term. And,
\textit{vice versa}, if the supersymmetry transformations of the 3-forms
contain chiralinos, then the Nambu-Goto term must contain a function of the
complex scalar fields to cancel it.

In Ref.~\cite{Hartong:2009az} consistent on-shell supersymmetry transformation
rules for two 3-forms transforming into the gravitino were found. The on-shell
condition is quite different for both in spite of the fact that their
supersymmetry transformations are the real and imaginary part of the same
expression: one of them says that the field strength is the dual of the scalar
potential (the part that depends on the superpotential) and the other says
that the field strength must vanish identically. The interpretation is that
one of them is associated to a coupling constant/deformation parameter of the
theory and the other is not.

This asymmetry is a bit surprising and a first hint that it is simply the
result of an asymmetrical description of the theory comes from the observation
(related to a similar observation made in Section~\ref{sec:N1d4}) that, if we
replace everywhere (except in the supersymmetry transformations of these
3-forms) $\mathcal{L}$ by $i\mathcal{L}$, a redefinition that does not change
the bosonic Lagrangian, the roles of the two 3-forms and their on-shell
conditions are interchanged. Replacing $\mathcal{L}$ by
$\frac{1}{\sqrt{2}}(1+i)\mathcal{L}$ instead, we find that the two 3-forms
play entirely analogous roles.

This suggests that the coupling constant/deformation parameter associated to
the superpotential is complex and the 3-forms are associated to the its real
and imaginary parts.

To make this relation explicit

\begin{enumerate}
\item We replace everywhere $\mathcal{L}$ by $(g^{1}+ig^{2})\mathcal{L}$ where
  $g^{1}$ and $g^{2}$ will be the two coupling constants. 

\item This means that, in the bosonic Lagrangian, the part of the scalar
  potential that depends on the superpotential is rescaled by a factor
  $(g^{1})^{2}+(g^{2})^{2}$. In the supersymmetry transformation rules, only
  those of the gravitino and chiralino are modified by this
  rescaling. Finally, the only parameter in the local supersmmetry algebra in
  Ref.~\cite{Hartong:2009az} that is modified is $\Lambda_{\mu\nu}$ which
  becomes

  \begin{equation}
  \Lambda_{\mu\nu} = -C_{\mu\nu\rho}\xi^{\rho} 
+2\Re{\rm e}\, [(g^{1}+ig^{2})\mathcal{L}\phi_{\mu\nu}]\, ,
\hspace{1cm}
\phi_{\mu\nu} = \bar{\epsilon}^{*}\gamma_{\mu\nu}\eta^{*}\, ,
  \end{equation}

\noindent
where $C_{\mu\nu\rho}$ is a 3-form to be determined that appears in the
2-forms field strengths.
 
\item We observe that there is a complex 3-form
  $\mathcal{C}_{\mu\nu\rho}=C^{1}{}_{\mu\nu\rho}+iC^{2}{}_{\mu\nu\rho}$ with
  supersymmetry transformation rules

\begin{equation}
\label{eq:3formansatz}
\delta_{\epsilon} \mathcal{C}_{\mu\nu\rho} 
=
12i \mathcal{L}^{*} \bar{\epsilon}\gamma_{[\mu\nu}\psi_{\rho]} 
+2\mathcal{D}_{i^{*}}  \mathcal{L}^{*}
\bar{\epsilon}\gamma_{\mu\nu\rho}\chi^{*i^{*}}\, ,    
\end{equation}

\noindent
such that

\begin{equation}
[\delta_{\eta},\delta_{\epsilon}] \mathcal{C}_{\mu\nu\rho} 
=
\pounds_{\xi}\mathcal{C}_{\mu\nu\rho}
+2\partial_{[\mu}\tilde{\Lambda}_{\nu\rho]}
+[\mathcal{G} - (g^{1}+ig^{2})\star V(\mathcal{L})]_{\mu\nu\rho\sigma}
\xi^{\sigma}\, ,
\end{equation}

\noindent
where

\begin{equation}
\mathcal{G} 
\equiv 
d\mathcal{C}\, ,  
\hspace{1cm}
\tilde{\Lambda}_{\mu\nu} 
\equiv
-\mathcal{C}_{\mu\nu\rho}\xi^{\rho}
+4i \frac{(g^{1}+ig^{2})}{|g^{1}+ig^{2}|^{2}}
[(g^{1}-ig^{2})\mathcal{L}^{*}\phi^{*}_{\mu\nu}]\, .
\end{equation}

\item The supersymmetry algebra closes if 

  \begin{equation}
   \mathcal{G} = (g^{1}+ig^{2})\star V(\mathcal{L}) \, ,
  \end{equation}

\noindent
which we can rewrite in components

\begin{equation}
G^{i} = \tfrac{1}{2}\star \frac{\partial V}{\partial g^{i}}\, ,
\hspace{1cm}
G^{i}\equiv dC^{i}\, ,\,\,\, i=1,2\, ,
\end{equation}

\noindent
so each of the real 2-forms is associated to a real coupling
constant/deformation parameter, as expected on general grounds.

\item Comparing the gauge parameter 2-form $\tilde{\Lambda}_{\mu\nu}$ with
  $\Lambda_{\mu\nu}$ above, we conclude that the 3-form that appears in the
  2-form field strengths is

\begin{equation}
C_{\mu\nu\rho} =\tfrac{1}{2}\Im{\rm m}\, [(g^{1}-ig^{2})\mathcal{C}] =
\tfrac{1}{2}(g^{1}C^{2}_{\mu\nu\rho} -g^{2}C^{1}_{\mu\nu\rho})\, ,
\end{equation}

\noindent
while

\begin{equation}
C^{\prime}_{\mu\nu\rho} =\tfrac{1}{2}\Re{\rm e}\, [(g^{1}-ig^{2})\mathcal{C}] =
\tfrac{1}{2}[g^{1}C^{1}_{\mu\nu\rho} +g^{2}C^{2}_{\mu\nu\rho}]\, ,
\end{equation}

\noindent
decouples. For a single coupling constant ($g^{2}=0$) taking the value
$g^{1}=1$ we recover the two 3-forms of Ref.~\cite{Hartong:2009az}:

\begin{equation}
C= \tfrac{1}{2}C^{2}\, ,
\hspace{1cm}
C^{\prime}= \tfrac{1}{2}C^{1}\, .  
\end{equation}

\item Finally, we can construct a supersymmetric worldvolume action 

  \begin{equation}
\label{eq:q12dwaction}
 S_{\mathrm{DW}}= |q^{1}+iq^{2}|\int d^{3}\xi \left\{|\mathcal{L}| \sqrt{|g_{(3)}|} 
+\tfrac{1}{8\cdot 3!} q^{i}\epsilon^{mnp} C^{i}{}_{mnp}\right\}\, ,
\end{equation}

\noindent
where $|g_{3}|$ the determinant of the pullback $g_{(3)\, mn}$ of the
spacetime metric to the 3-dimensional worldvolume and $C_{mnp}$ is the
pullback of the 3-form to the worldvolume:

\begin{equation}
g_{(3)\, mn}  
\equiv 
g_{\mu\nu}\frac{\partial X^{\mu}}{\partial \xi^{m}}
\frac{\partial X^{\nu}}{\partial \xi^{n}}\, ,
\hspace{1.5cm}
C_{mnp} \equiv
C_{\mu\nu\rho}
\frac{\partial X^{\mu}}{\partial \xi^{m}}\frac{\partial X^{\nu}}{\partial
  \xi^{n}}
\frac{\partial X^{\rho}}{\partial \xi^{p}}\, .
\end{equation}

\noindent
It is convenient to work in the static gauge in which we identify the
worldvolume coordinates $\xi^{m}$ with the first three spacetime coordinates
$X^{m}$, so

\begin{equation}
\frac{\partial X^{\mu}}{\partial \xi^{m}} =\delta^{\mu}{}_{m}\, ,  
\end{equation}

\noindent
and 

\begin{equation}
g_{(3)\, mn}
=
g_{\underline{m}\underline{n}}\, ,
\hspace{1cm}
C_{mnp}  
= 
C_{\underline{m}\underline{n}\underline{p}}\, .  
\end{equation}

It is then straightforward to see that the above action is invariant to lowest
order in fermions under the supersymmetry transformations
Eqs.~(\ref{eq:susytranseN1}, \ref{eq:susytransZN1}) and (\ref{eq:3formansatz})
iff  the spinors satisfy the condition

\begin{equation}
e^{-i(\alpha+q)/2}\epsilon 
+\gamma^{012} (e^{-i(\alpha+q)/2}\epsilon)^{*} =0\, ,
\hspace{1cm}
e^{iq}\equiv \frac{q^{1}+iq^{2}}{|q^{1}+iq^{2}|}\, ,\,\,\,
e^{i\alpha} \equiv \frac{\mathcal{L}}{|\mathcal{L}|}\, .
\end{equation}

\noindent
which generalizes Eq.~(\ref{eq:dwsusyprojector}).

\end{enumerate}

\subsection{Sourceful supersymmetric domain-wall solutions}
\label{sec:Shebang}

Now we are going to couple the action Eq.~(\ref{eq:q12dwaction}) found in the
previous section to the bulk $N=,d=4$ action to use it as a source term for
domain-wall solutions.  We will only consider the $q^{2}=C^{2}=0$ case for the
sake of simplicity, renaming $C\equiv C^{1}/2$.

However, we cannot couple it to the bulk supergravity action,
Eq.~(\ref{eq:actionN1ungauged}), by simply adding them up because the 3-form
only occurs in the source. As discussed before, we must promote the coupling
constant $g\equiv g^{1}$ to a scalar field $g(x)$ and add to the bulk
supergravity action a Lagrange-multiplier term containing the 3-form as to
enforce the constancy of $g$. Thus, we are led to consider the bulk
supergravity action,

\begin{equation}
\label{eq:bulk}
 S_{\rm bulk}  =  \frac{1}{\kappa^{2}}{\displaystyle\int} d^{4}x \sqrt{|g|}
\left[R +2\mathcal{G}_{ij^{*}}\partial_{\mu}Z^{i}
\partial^{\mu}Z^{*\, j^{*}} 
-g^{2}(x)V(Z,Z^{*}) -{\textstyle\frac{1}{
  3\sqrt{|g|}}} \epsilon^{\mu\nu\rho\sigma}\partial_{\mu}g(x) C_{\nu\rho\sigma}
\right]\, ,
\end{equation}

\noindent
and the brane-source action 

\begin{equation}
\label{eq:dwaction-2}
 S_{\mathrm{DW}}=-\int d^{4}x f(y) \left\{|\mathcal{L}| \sqrt{|g_{(3)}|} 
\pm\tfrac{1}{4!}\epsilon^{mnp} C_{\underline{m}\underline{n}\underline{p}}\right\}\, ,
\end{equation}

\noindent
where $f(y)$ is a distribution-function of domain walls with a common
transverse direction parametrized by the coordinate $x^{3}\equiv y$. For
instance we could take $f(y)=\delta (y-y_{0})$ as to treat the case of a
single infinitely-thin domain wall placed at $y=y_{0}$. Having a source term
for these domain walls we can do away with the Israel junction conditions
\cite{Israel:1966rt}. It is not clear if more complicated, continuous
distributions can be derived from the single brane effective action
Eq.~(\ref{eq:q12dwaction}), but we will use them as toy models.

The equations of motion that follow from the total action $S\equiv S_{\rm
  bulk}+S_{\rm DW}$ are 

\begin{eqnarray}
\mathcal{E}_{g}^{\mu\nu}  
& = &
-\frac{\kappa^{2}}{2}f(y) |\mathcal{L}|
\frac{\sqrt{|g_{(3)}|}}{\sqrt{|g|}}
g_{(3)}^{mn}\delta_{m}{}^{\mu}\delta_{n}{}^{\nu}\, , 
\\
& & \nonumber \\
\mathcal{G}^{ij^{*}}\mathcal{E}_{g\, i^{*}} 
& = &
-\frac{\kappa^{2}}{8}f(y)
\frac{\sqrt{|g_{(3)}|}}{\sqrt{|g|}}
e^{i\alpha}\mathcal{N}^{i}\, , 
\\
& & \nonumber \\
\epsilon^{\mu\nu\rho\sigma}\partial_{\sigma}g(x)
& = & 
\pm
\frac{\kappa^{2}}{8}f(y) \epsilon^{mnp}
\delta_{m}{}^{\mu}\delta_{n}{}^{\nu}\delta_{p}{}^{\rho}\, ,
\\
& & \nonumber \\
\epsilon^{\mu\nu\rho\sigma}\partial_{\mu}C_{\nu\rho\sigma}
& = & 
6\ g(x)\; V(Z,Z^{*})\, ,
\end{eqnarray}

\noindent
where $\mathcal{E}_{g}^{\mu\nu}$ and $\mathcal{E}_{g\, i^{*}}$ are the
Einstein and scalar equations of motion defined in Eqs.~(\ref{eq:Emn}) and
(\ref{eq:Ei}) after the introduction of $g(x)$. Observe that the
Lagrange-multiplier term is topological and independent of the scalars and,
furthermore, does not modify the Einstein nor the scalar equation of motion.

The third equation is that of the 3-form and is solved if $g$ is a function of
$y$ satisfying

\begin{equation}
\label{eq:gequation}
\partial_{\underline{y}}g = \pm \tfrac{1}{8}\kappa^{2}f(y)\, .
\end{equation}

\noindent
The function $g(y)$ will have step-like discontinuities at the locations of
the domain walls, in the case they are infinetely thin. 

The fourth equation is that of the scalar $g(x)$ and, as required, simply states that the
3-form is the dual of the scalar potential.

It can be checked that the Einstein and scalar equations of motion are
identically satisfied if the metric function $H(y)$ and the scalars $Z^{i}(y)$
satisfy the following modified \textit{sourceful flow equations}

\begin{eqnarray}
\label{eq:sourcefulflowZi}
\partial_{\underline{y}}Z^{i}
& = & 
\pm g(y) e^{i\alpha}\mathcal{N}^{i}H^{1/2}\, ,
\\
& & \nonumber \\
\label{eq:sourcefulflowH}
\partial_{\underline{y}}H^{-1/2} 
& = & 
\pm 2 g(y) |\mathcal{L}|\, .  
\end{eqnarray}

These equations can be derived following the procedure of
Section~\ref{sec-SSDWS} using the modified fermion supersymmetry
transformation rules

\begin{eqnarray}
\label{eq:modifiedgravisusyruleN1}
\delta_{\epsilon}\psi_{\mu} & = & 
\mathcal{D}_{\mu}\epsilon +ig(x)\mathcal{L}\gamma_{\mu}\epsilon^{*}
\, ,
\\
& & \nonumber \\
\label{eq:modifieddilasusyruleN1}
\delta_{\epsilon}\chi^{i} & = & 
i\not\!\partial Z^{i}\epsilon^{*}+g(x)\mathcal{N}^{i}\epsilon\, .
\end{eqnarray}

It should be clear that a fully supersymmetric formulation of $N=1,d=4$
supergravity including all higher-rank forms is necessary in order to make
sense of these modifications.  Likewise, it should also be clear that these
modifications have a non-trivial effect on the source-free solutions, as will
be illustrated by means of a simple example.

\subsection{A simple example}
\label{sec:Example}

Let us consider model of $N=1,d=4$ supergravity coupled to a single chiral
multiplet defined by the K\"ahler potential and superpotential

\begin{equation}
\mathcal{K}= |Z|^{2}\, ,
\hspace{1cm} 
W=1\, ,\,\,\,
(\mathcal{L} = e^{|Z|^{2}/2})\, . 
\end{equation}

The fermion shift $\mathcal{N}^{Z}$ is given by

\begin{equation}
\mathcal{N}^{Z} = 2\ Z\ e^{|Z|^{2}/2}\, , 
\end{equation}
and the scalar potential is readily seen to be 

\begin{equation}
V = -8(3-\rho^{2})e^{\rho^{2}/2}\, ,  
\end{equation}
\begin{figure}
\centering
\includegraphics[height=3cm]{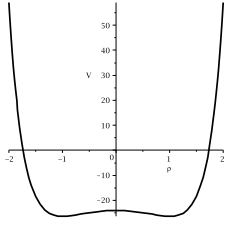}
\caption{\label{fig:VThin} The potential $V$ as a function of $\rho$.}
\end{figure}

\noindent
where we have defined $Z\equiv \rho e^{i\beta}$, so that $\rho$ is a radial
coordinate.  As one can see from the plot of this potential in figure
(\ref{fig:VThin}), this potential has a degenerate (local) maximum at $\rho=0$
and a degenerate (absolute) minimum at $\rho=+1$ and takes negative values at
both of them. At these extrema the values of the potential are

\begin{equation}
V(0) \ =\ -24\, ,
\hspace{1cm}
V(1)\ =\ -16\ \sqrt{e} \, .  
\end{equation}

\noindent
The absolute value of these numbers is not relevant, as $V$ is 
multiplied by $g^{2}(y)$, a factor which is determined by the sources.

The sourceful flow equation (\ref{eq:sourcefulflowZi}) implies that the
argument of $Z$, $\beta$, is constant. Then, Eqs.~(\ref{eq:sourcefulflowZi})
and (\ref{eq:sourcefulflowH}) take the form

\begin{eqnarray}
\label{eq:sourcefulflowZexample}
\partial_{\underline{y}}\rho
& = & 
\pm 2 g(y) \rho e^{\rho^{2}/2}  H^{1/2}\, ,
\\
& & \nonumber \\
\partial_{\underline{y}}H^{-1/2} 
& = & 
\pm 2 g(y) e^{\rho^{2}/2}\, .  
\end{eqnarray}

\noindent
The first equation implies that, in a region in which $\rho$ (and, hence, $Z$)
is constant, the product $g(y)\rho$ must vanish. Thus, for constant $g(y)=g$
(i.e.~in absence of sources) $\rho=0$ provides a solution with
$AdS_{4}$ metric 

\begin{equation}
H= \frac{1}{4g^{2}y^{2}}\, .  
\end{equation}

\noindent
$\rho=1$, however, can only be a solution if $g=0$, in which case we have a
Minkowski spacetime. This may look strange at first sight since we found that
the value of the potential for $\rho=1$ is $V(1) = -16\sqrt{e}$ and we might
have expected an $AdS_{4}$ solution. Of course, such an $AdS_{4}$ solution
exists, but it is not supersymmetric and moreover does not satisfy the
sourceless flow equations.

For non-constant $\rho$ we can combine the two sourceful flow equations to find

\begin{equation}
\partial_{\underline{y}} \log{\rho}= \partial_{\underline{y}} \log{H^{-1/2}}\, ,  
\end{equation}

\noindent
so that

\begin{equation}
H = c/\rho^{2}\, ,  
\end{equation}

\noindent
for some real and positive integration constant $c$. Substituting this
expression of $H$ into Eq.~(\ref{eq:sourcefulflowZexample}) we get

\begin{equation}
\label{eq:firstsolution}
\rho
=
\sqrt{2}\ \mathrm{Erf}^{-1}\left[G(y)\right] \, , 
\end{equation}

\noindent
where $\mathrm{Erf}^{-1}$ is the inverse of the normalized error function\footnote{
The normalized error function is defined by
\begin{equation}
\mathrm{Erf}(x) \equiv \frac{2}{\sqrt{\pi}}\int_{0}^{x} e^{-u^{2}}du 
\ =\ -\mathrm{Erf}(-x) \, ,
\end{equation}
and grows monotonically between $\mathrm{Erf}(-\infty)=-1$ and
$\mathrm{Erf}(\infty)=1$.  Around the points $x=0,1$ it admits the expansions
\begin{equation}
  \begin{array}{rcl}
   \mathrm{Erf}(x) & = & \frac{2}{\sqrt{\pi}}\left\{x -\frac{x^{3}}{3}+\ldots
   \right\}\, ,\\
& & \\ 
   \mathrm{Erf}(x) & = & 1 -\frac{e^{-x^{2}}}{\sqrt{\pi}x}\left\{1 +\ldots \right\}\, .\\
  \end{array}
\end{equation}
}
\noindent
and

\begin{equation}
G(y)\; \equiv\;  \pm \sqrt{\frac{8c}{\pi}}\ \int g(y) dy \; +\; d\; ,  
\end{equation}

\noindent
where $d$ is another integration constant.  Observe that for $\rho$ to be a
well-defined radial coordinate, {\em i.e.\/} $\rho \geq 0$, $G(y)$ has to take
values inside the interval $[0,1)$, where we exclude the value $g(Y)=1$ as it
corresponds to $\rho =\infty$.

To solve completely our problem we must define a domain-wall source
distribution function $f(y)$ to determine $g(y)$ by means of
Eq.~(\ref{eq:gequation}). Let us consider, first, a single, infinitely thin
domain wall of positive tension $q>0$ placed at $y=y_{0}$, described by the
distribution function

\begin{equation}
f(y)\; =\; q\ \delta (y-y_{0})\, .  
\end{equation}

\noindent
The corresponding local coupling constant $g(y)$ is\footnote{$\theta(x)$ is
  the Heaviside $\theta$-function which is $\theta(x) =1$ for $x>0$ and zero
  otherwise.}

\begin{equation}
g(y) \; =\;  \pm \frac{\kappa^{2}q}{16}\ [\theta (y-y_{0}) -\theta(y_{0}-y)]
     \; =\;  \pm \frac{\kappa^{2}q}{16}\ \mathrm{sgn}(y-y_{0})\, , 
\end{equation}

\noindent
where $\mathrm{sgn}$ is the signum function. The above can be trivially
integrated to give

\begin{equation}
  G(y) \; =\;  q\kappa^{2}\ \sqrt{\frac{c}{32\pi}}\ |y-y_{0}|\; +\; d\, .
\end{equation}

As it stands, the range of $G$ is unbounded and is therefore not completely
contained in the domain of $\mathrm{Erf}^{-1}$, whence the solution
(\ref{eq:firstsolution}) is not well-defined.  A possible way of obtaining a
$G$ whose range is contained in the domain is by introducing a second,
parallel, domain wall with opposite tension and charge at a different point
($y=-y_{0}$ with $y_{0}>0$ for simplicity).  This means that
%

\begin{equation}\label{eq:DoubleThin}
  \begin{array}{rcl}
f(y) & = & q\delta (y-y_{0})-q\delta (y+y_{0})\, ,  \\
& & \\
g(y) & = & 
\pm {\displaystyle\frac{\kappa^{2}q}{16}}[\theta (y-y_{0}) -\theta(y_{0}-y)
-\theta (y+y_{0}) +\theta(-y_{0}-y)]\, ,\\ 
& & \\
G(y) & = & 
{\displaystyle\sqrt{\frac{c}{32\pi}}}\kappa^{2}q
\left(|y-y_{0}|-|y+y_{0}|\right)+d\, .\\
\end{array}
\end{equation}

\noindent
\begin{figure}
\centering
\includegraphics[height=3cm]{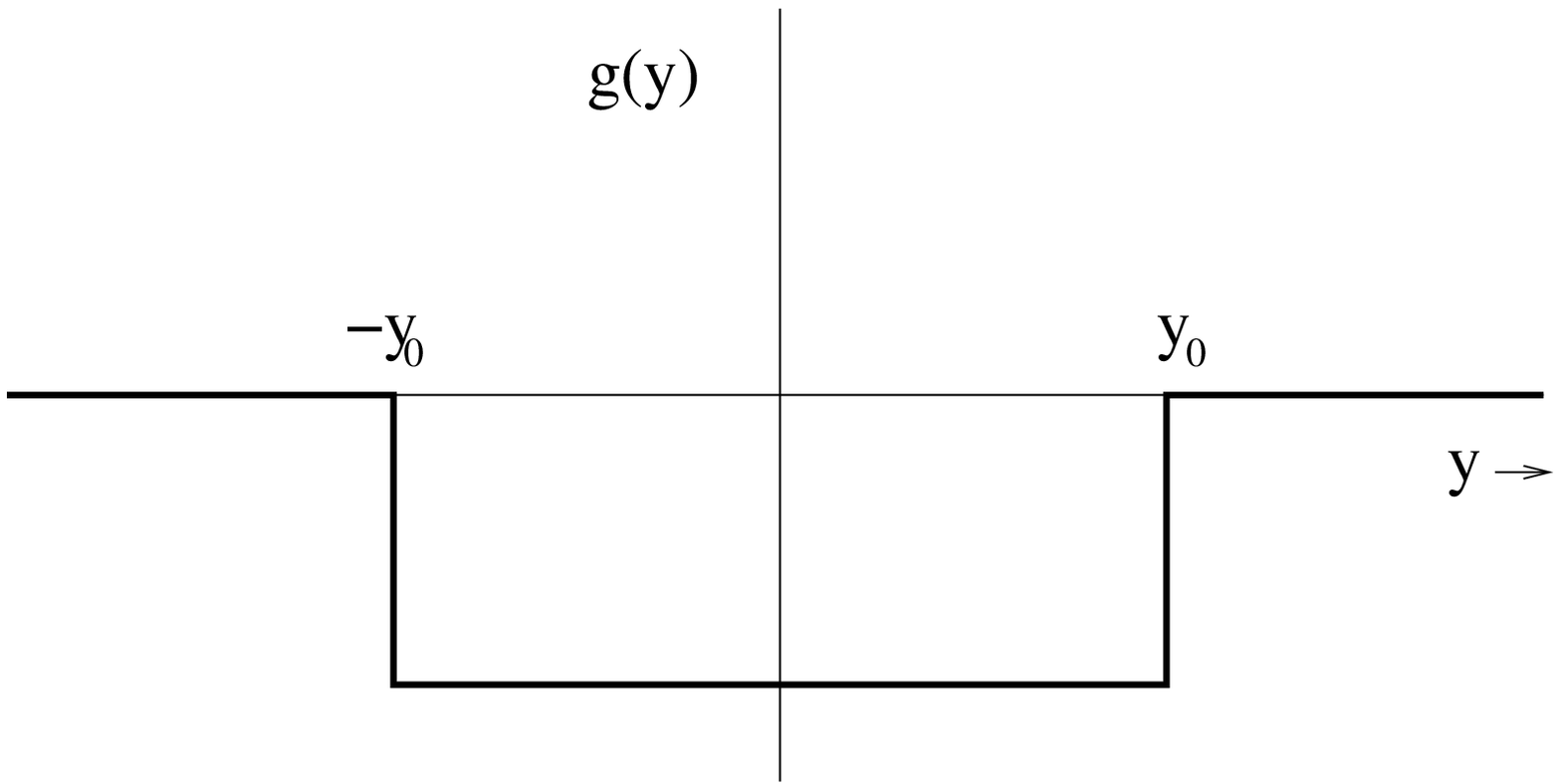}
\hspace{2cm}
\includegraphics[height=3cm]{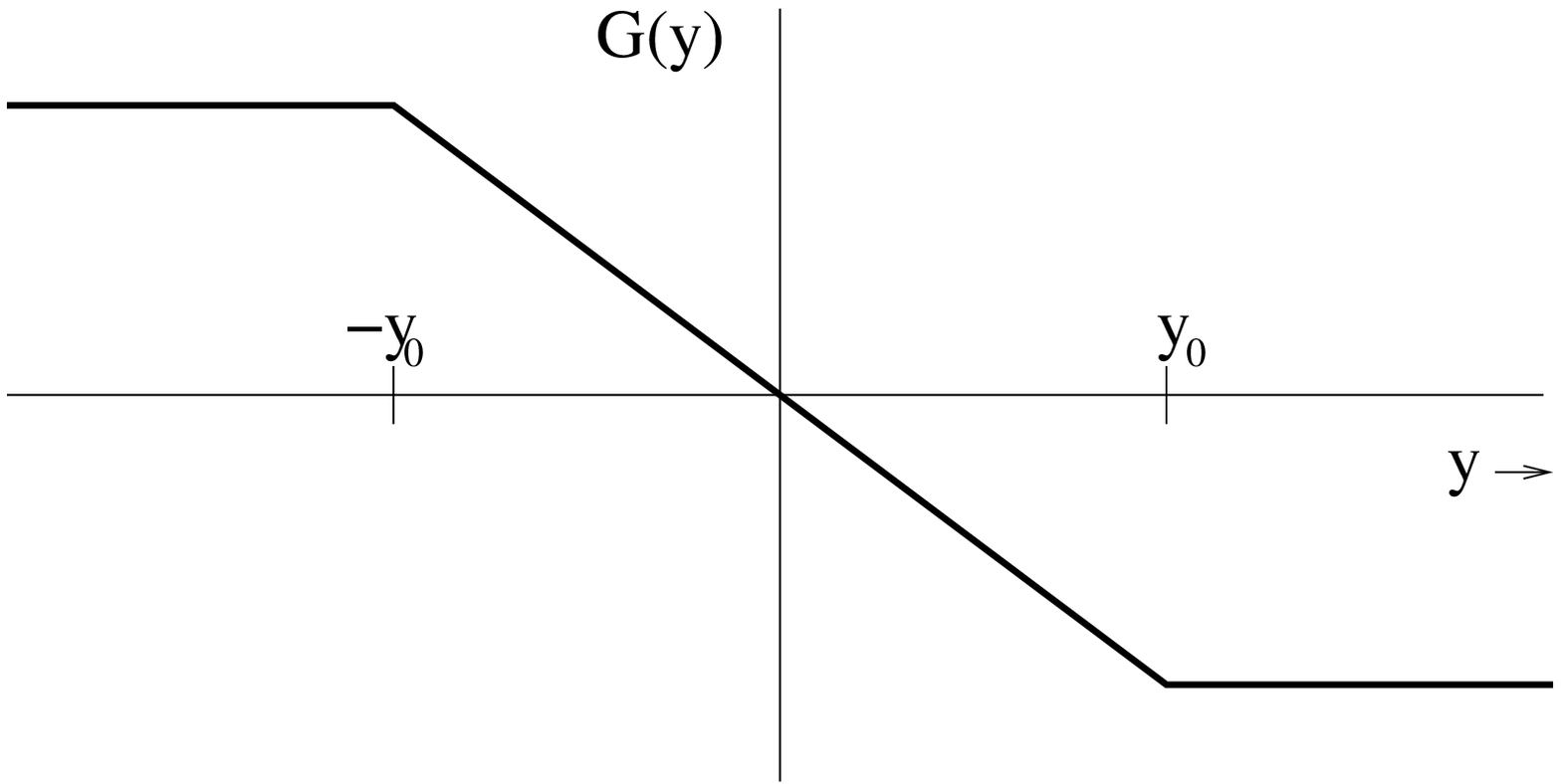}
\caption{\label{fig:gandGThin} The profiles of the functions $g(y)$ and $G(y)$ as given in 
                               Eqs.~(\ref{eq:DoubleThin}).
}
\end{figure}
In other words: on the interval $[-y_{0},y_{0}]$, $g(y)$ takes on the constant
value $\mp {\displaystyle\frac{\kappa^{2}q}{8}}$ and vanishes identically
outside said interval; this implies that outside the interval the scalars $Z$
and $Z^{*}$ vanish, whence the the spacetime for $|y|> y_{0}$ is Minkowski.
The function $G$ is constant outside the interval and on the interval it
interpolates linearly between these two constant values which we will
denominate $G(-\infty )$ and $G(\infty )$.  These asymptotic values are given
by
\begin{equation}
G(\mp\infty)\; =\; \pm q\kappa^{2}\ {\displaystyle\sqrt{\frac{c}{8\pi}}}\ y_{0}\, +\, d\, .
\end{equation}


In order to create a well-behaved solution we must choose the integration
constants judiciously: as $G(y)$ decreases on the interval $[-y_{0},y_{0}]$
and $\mathrm{Erf}^{-1}$ is a monotonic function, whence $\rho(y)$ also
decreases on the interval, it is natural to choose $d$ so as to make
$G(+\infty)=G(+y_{0})=0$.  This implies that $\rho(+y_{0})=0$ and we can make
$\rho$ continous across the domain wall located at $y=y_{0}$, by choosing
$\rho(+\infty)=0$.  This means taking

\begin{equation}
d\; =\; q\kappa^{2}\ {\displaystyle\sqrt{\frac{c}{8\pi}}}\ y_{0}\; .  
\end{equation}

It is interesting to study how the solution approaches the point $y=y_{0}$
from the interior of the $g(y)\neq 0$ region. The scalar field approaches zero
as

\begin{equation}
\rho\; \sim\; \tfrac{q\kappa^{2}}{4}\ \sqrt{c}\ (y_{0}-y)\, ,  
\end{equation}

\noindent
so the metric approaches that of $AdS_{4}$

\begin{equation}
H \sim \frac{R^{2}}{(y_{0}-y)^{2}}\, ,
\hspace{1cm}
R= \frac{4}{\kappa^{2}q}\, .  
\end{equation}

\noindent
The limit $y=y_{0}$ is actually at an infinite proper distance
in spacelike directions.

Let us now consider the other end of the $g(y)\neq 0$ region, $y=-y_{0}$,
where $G(y)$ reaches the value $G(-y_{0})=
\sqrt{\frac{c}{2\pi}}\kappa^{2}qy_{0}= G(-\infty)$, which can be tuned by
moving the domain-wall sources ($y_{0}$). Thus value has to be smaller or
equal than 1 in order to have a well-defined solution.

The cases $G(-y_{0})=1$ and $G(-y_{0})< 1$ are very different: if $G(-y_{0})<
1$ then $\rho(-y_{0})$ is finite and we can choose the constant value of
$\rho$ in the $y<-y_{0}$ $\rho(-\infty)=\rho(-y_{0})$ to have continuity.
$\rho$ approaches $y=-y_{0}$ from the interior of the $g(y)\neq 0$ region
as

\begin{equation}
\rho\sim  -\sqrt{\frac{c}{2\pi}} 
\frac{\kappa^{2}q}{\mathrm{Erf}^{\, \prime}[\rho(-\infty)/\sqrt{2}]}
(y+y_{0})\, ,
\end{equation}

\noindent
so the metric approaches another $AdS_{4}$ region and we can do without the
exterior $y< -y_{0}$ region. The solution we have obtained smoothly
interpolates between two $AdS_{4}$ regions one of which (the $\rho=0$ one)
corresponds to a supersymmetric vacuum of the bulk-theory.

Let us point out the the effective potential $g^{2}(y)V$ evaluated on this
solution can become positive near $y=y_{0}$ when $G(-\infty
)>\mathrm{Erf}(\sqrt{3/2})$.

The two infinitely thin domain-wall sources set-up may be understood as an
approximation to the following configuration with domain-wall sources of
{\em finite thickness} described, for instance,  by 

\begin{equation}
  \begin{array}{rcl}
f(y) & = & qy e^{-y^{2}}\, ,  \\
& & \\
g(y) & = & 
\mp {\displaystyle\frac{\kappa^{2}q}{16}} e^{-y^{2}}\, ,\\ 
& & \\
G(y) & = & 
-{\displaystyle\frac{q\kappa^{2}\ \sqrt{2c}}{16}}
\mathrm{Erf}(y)+d\, .\\
\end{array}
\end{equation}

\noindent
Observe that the local coupling constant $g(y)$ vanishes only asymptotically.
\par
\begin{figure}
\centering
\includegraphics[width=9cm]{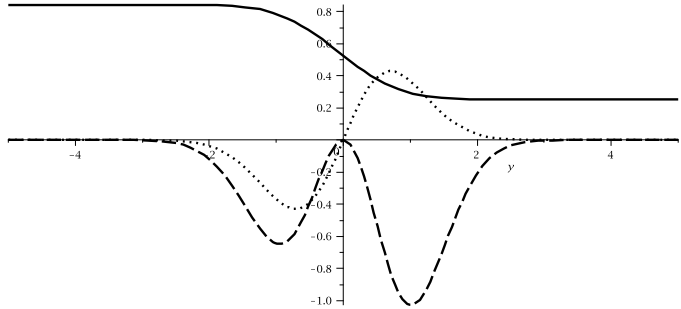}
\caption{\label{fig:FinThick} The various profiles of the functions ocurring
  in the solution of the thick domain wall for the regular case: the dotted
  line corresponds to the chosen source, $f(y)$, the solid line to the
  corresponding scalar field $\rho (y)$, and the dashed line corresponds to
  the effective potential as seen by the solution, {\it i.e.\/}
  $g^{2}(y)V$. Observe that the degeneracy of minima of the potential is
  broken by the local coupling constant.}
\end{figure}

This source distribution will lead to a well-defined scalar field $\rho$ if $0
\leq d- q\kappa^{2} \sqrt{2c}/16 < 1$; if we choose $d=
q\kappa^{2}\sqrt{2c}/16$, so that $G(y)= {\displaystyle\frac{q\kappa^{2}
    \sqrt{2c}}{16}}\ [1-\mathrm{Erf}(y)]$, $\rho$ will vanish asymptotically
as $\rho\sim e^{-y^{2}}/y$ when $y\rightarrow +\infty$ and the metric will
diverge as $H\sim \rho^{-2}$ in that limit. If we choose $d > q\kappa^{2}
\sqrt{2c}/16$, $\rho$ will asymptote to a constant value $\rho \sim
\rho(+\infty)$ and the metric will be asymptotically flat.  The same happens
in the $y\rightarrow -\infty$ limit if we take $d + q\kappa^{2} \sqrt{2c}/16<
1$, which can be arranged by an adequate choice of $c$; $\rho$ will, however,
diverge in that limit if we choose $d + q\kappa^{2}\sqrt{2c}/16=1$.  As shown
in figure (\ref{fig:FinThick}), the fact that the asymptotically non-diverging
solutions interpolate between asymptotic Minkowskian spaces is due to the fact
that the effective potential as seen by the solution, {\it i.e.\/} $g^{2}V$,
vanishes asymptotically.

This asymptotic behavior is qualitatively similar to that of the
infinitely-thin sources case. On the other hand, if we expand the solution
around any finite value of $y$ we will find that the metric is locally
$AdS_{4}$, as happens in the infinitely-thin sources case on the interval
$[-y_{0},+y_{0}]$.

As we have seen, the introduction of sources, which forces the introduction of
local coupling constants, modifies the domain-wall solutions dramatically.

\section{Supersymmetric instantons}
\label{sec:instantons}

\subsection{General sourceless and sourceful instanton solutions}
\label{sec:GenInst}

We are interested in (multi-) instanton solutions of $N=1,d=4$ supergravity
generalizing the stringy D-instanton of Ref.~\cite{Gibbons:1995vg}, i.e.~with
flat Euclidean metric and unbroken supersymmetry.

It is convenient to start by considering the general case of a
$d_{\sigma}$-dimensional $\sigma$-model with real coordinates $\phi^{i}$
$i=1,\ldots, d_{\sigma}$ and metric $\mathcal{G}_{ij}$ coupled to gravity in
$d$ Euclidean dimensions with action (up to boundary terms)

\begin{equation}
\label{eq:genericaction}
S_{\rm bulk} = \int d^{d}x\sqrt{|g|}\{R +
\textstyle{\frac{1}{2}}\mathcal{G}_{ij}\partial_{\mu}\phi^{i}\partial^{\mu}\phi^{j}
\}\, ,  
\end{equation}

\noindent
and equations of motion

\begin{eqnarray}
R_{\mu\nu} 
+\textstyle{\frac{1}{2}}\mathcal{G}_{ij}\partial_{\mu}\phi^{i}\partial_{\nu}\phi^{j}  
& = & 
0\, ,
\\
& & \nonumber \\
\label{eq:scalareom}
\nabla^{2}\phi^{i} +\Gamma_{jk}{}^{i}\partial_{\mu}\phi^{j} \partial^{\mu}\phi^{k}
& = & 
0\, .
\end{eqnarray}

The requirement that the instanton solution has flat metric
$g_{\mu\nu}=\delta_{\mu\nu}$ so $R_{\mu\nu}=0$ in the Einstein equation
implies

\begin{equation}
\label{eq:nullcondition1}
\mathcal{G}_{ij}\partial_{\mu}\phi^{i}\partial_{\nu}\phi^{j}  =0\, .  
\end{equation}

Thus, the kind of instantons we are looking for only exists when the Euclidean
action is not positive-definite.

Now, following Ref.~\cite{Neugebauer:1969wr}, we assume that the scalars in the
instanton solution depend on $n$ independent functions of the $d$ spatial
coordinates $\sigma^{a}(x)\, ,\,\,\,a=1,\ldots, n$. The equations of motion
(\ref{eq:nullcondition1}) and (\ref{eq:scalareom}) become, respectively,

\begin{eqnarray}
\mathcal{G}_{ij}\partial_{a}\phi^{i}\partial_{b}\phi^{j} 
\partial_{\mu}\sigma^{a}\partial_{\nu}\sigma^{b} 
& = & 
0\, ,  
\\
& & \nonumber \\
\nabla^{2}\sigma^{a}\partial_{a}\phi^{i} 
+\left\{\partial_{a}\partial_{b}\phi^{i} 
+\Gamma_{jk}{}^{i}\partial_{a}\phi^{j} \partial_{b}\phi^{k} \right\}
\partial_{\mu}\sigma^{a}\partial_{\mu}\sigma^{b}
& = & 
0\, ,  
\end{eqnarray}

\noindent
where $\partial_{a}\equiv \partial/\partial \sigma^{a}$. If we assume that the
$\sigma^{a}$ are harmonic functions

\begin{equation}
\nabla^{2}\sigma^{a} =0\, ,  
\end{equation}

\noindent
(up to singular terms in the r.h.s.~that will be dealt with when we consider
the sources), then these solutions are satisfied if we find an $n$-dimensional
hypersurface $\phi^{i}(\sigma)$ with $n$ null and mutually orthogonal tangent
vectors $\partial_{a}\phi^{i}(\sigma)$ satisfying the equations

\begin{equation}
\label{eq:multigeodesic}
\partial_{a}\partial_{b}\phi^{i} 
+\Gamma_{jk}{}^{i}\partial_{a}\phi^{j} \partial_{b}\phi^{k}=0\, ,
\end{equation}

\noindent
which imply (for each value of $a=b$) that each of the $n$ coordinate curves
obtained by taking a constant value of all but one of the $\sigma^{a}$s are
geodesics in the scalar target space.

These equations can be solved by finding $n$ null and mutually orthogonal
vector fields $N_{a}{}^{i}(\phi)$ in the scalar target space satisfying the
equations

\begin{equation}
\label{eq:paralleltransport}
N_{(a|}{}^{j}\nabla_{j}N_{|b)}{}^{i}=0\, ,
\hspace{1cm}
\forall\,  a,b=1,\ldots, n\, ,  
\end{equation}

\noindent
and then solving for $\phi(\sigma)$ the system of flow equations

\begin{equation}
N_{a}{}^{i} (\phi(\sigma)) = \partial_{a}\phi^{i}(\sigma)\, .  
\end{equation}

\noindent
This is possible only if the integrability conditions
$\partial_{[a}\partial_{b]}\phi^{i}=0$ are satisfied, i.e.~if the null vector
fields $N_{a}$ commute over the null hypersurface $\phi^{i}(\sigma)$:

\begin{equation}
\partial_{[a}N_{b]}{}^{i}
=
\partial_{[a}\phi^{j}\partial_{j}N_{b]}{}^{i} 
=   
\left. N_{[a}{}^{j}\partial_{j}N_{b]}{}^{i} \right|_{\phi(\sigma)}
=
\left. \tfrac{1}{2}[N_{a},N_{b}]^{i}\right|_{\phi(\sigma)}=0\, .
\end{equation}

\noindent
If the vector fields $N_{a}$ are Killing vectors this has further consequences
that we will not explore here.

Therefore, the problem of finding D-like instanton solutions can be reduced to
the problem of finding $n$ null, mutually orthogonal vector fields (which is
only possible if the dimensionality of the $\sigma$-model is greater or equal
than $2n$) satisfying the parallel-transport conditions
Eq.~(\ref{eq:paralleltransport}) in the space with $\sigma$-model metric
$\mathcal{G}_{ij}$. The instanton solutions are given by the integral surfaces
$\phi(\sigma)$ where the surface coordinates $\sigma$ satisfy the Laplace
equation in space.

It is always possible (but in no way necessary) to define coordinates
$\{\phi^{a+},\phi^{M}\}$ with $a=1,\ldots,n\, ,\,\,\, M=n+1,\ldots,
d_{\sigma}$ adapted to the null vector fields

\begin{equation}
  N_{a}{}^{i}\partial_{i} \equiv \partial_{a+}\, , 
\end{equation}

\noindent
so on the integral hypersurfaces $\phi^{a+}(\sigma)=\sigma^{a}$. We can also
define $n$ dual, mutually orthogonal 1-forms $L^{a}$, such that
$L^{a}{}_{i}L^{b}{}_{j}\mathcal{G}^{ij}=0$ and
$L^{a}{}_{i}N_{b}{}^{i}=\delta^{a}{}_{b}$, whose coordinate representation reads

\begin{equation}
  L^{a}{}_{i}d\phi^{i} = d\phi^{a+} +A^{a+}{}_{M}d\phi^{M}\, .   
\end{equation}

Let us now consider the introduction of sources for these instanton solutions:
as instantons have a 0-dimensional worldvolume, their effective actions are
just the value of some field at the location of the source. 
The most general action that we can write down, then, has the form

\begin{equation}
S_{\rm inst} =   \int d^{d}x f^{a}(x)\mathcal{F}_{a}(\phi)\, , 
\end{equation}

\noindent
where the $f^{a}(x)$ are some given distribution functions ($\sim
\delta^{(d)}(x)$ for a single instanton at the origin) and the
$\mathcal{F}_{a}(\phi)$s are linearly-independent functions of the scalar
fields whose properties will be determined by consistency (and, in due time,
by requiring supersymmetry).

The coupling of this action to the bulk action Eq.~(\ref{eq:genericaction})
modifies the equations of motion of the scalars Eq.~(\ref{eq:scalareom}) to 

\begin{equation}
\label{eq:scalareom2}
\nabla^{2}\phi^{i} +\Gamma_{jk}{}^{i}\partial_{\mu}\phi^{j} \partial^{\mu}\phi^{k}
 -\frac{f^{a}(x)}{\sqrt{g}}\mathcal{G}^{ij}\partial_{j}\mathcal{F}_{a}
=  
0\, .  
\end{equation}

Introducing the functions $\sigma^{a}$ this equation can be solved by solving
separately Eqs.~(\ref{eq:multigeodesic}) and

\begin{equation}
\nabla^{2}\sigma^{a} = \frac{f^{a}(x)}{\sqrt{g}} \, ,  
\end{equation}

\noindent
if we assume that 

\begin{equation}
\partial_{a}\phi^{i}=\mathcal{G}^{ij}\partial_{j}\mathcal{F}_{a}\, .   
\end{equation}

Observe that this condition implies that the 1-form dual to the null vectors
$N_{a}$ are exact:
$\partial_{a}\phi^{i}\mathcal{G}_{ij}d\phi^{j}=d\mathcal{F}_{a}$. If we want
to work in adapted coordinates $\phi^{a+}$ then we can use the functions
$\mathcal{F}_{a}(\phi)$ to define coordinates $\phi^{a-}\equiv
\mathcal{F}_{a}(\phi)$. Calling the remaining coordinates $\phi^{m}$ and using
these coordinates, the $\sigma$-model metric has to be of the Walker
type\footnote{In Ref.~\cite{art:walker1950}, A.G.~Walker asked the following
  question: Given a spacetime of dimension $n$ with a metric $g$, which admits
  $m$ ($2m\leq n$) independent null-vectors $N_{a}$ satsifying the propagation
  rule $\nabla\ N_{a}=\mathsf{A}_{a}{}^{b}\otimes N_{b}$, what is the
  canonical form of the metric? The answer is that any such metric can be
  written as in Eq.~(\ref{eq:adaptedmetric}).  }

\begin{equation}
\label{eq:adaptedmetric}
\mathcal{G}_{ij}d\phi^{i}d\phi^{j}
=
2d\phi^{a-}(d\phi^{a+} +A^{a+}{}_{b-}d\phi^{b-} +A^{a+}{}_{m}d\phi^{m})
+\mathcal{C}_{mn}d\phi^{m}d\phi^{n}\, ,
\end{equation}

\noindent
while the instanton source will take the form

\begin{equation}
\label{eq:adaptedinstantonsources}
S_{\rm inst} =   \int d^{d}x f^{a}(x)\phi^{a-}(x)\, . 
\end{equation}

In adapted coordinates the instanton solutions will have the form
$\phi^{a+}=\sigma^{a}\, ,\,\, \phi^{a-}=\mathrm{const.}\, ,\,\,
\phi^{m}=\mathrm{const.}$

At this point we must raise the question: given $n$ independent null-geodesics
of a metric $g$, under what conditions can they be embedded into
simultaneously-twistfree null-congruences? It is known from the literature
concerning the Penrose limit, that a null-geodesic on a Lorentzian space can
always be embedded into a twistfree null-congruence and the proof, see {\it
  e.g.\/} Ref.~\cite{Blau:2004yi}, can in all likelyhood be enhanced to the
case of one null-geodesic on a pseudo-Riemannian space, but it seems unikely
that such an embedding is always possible for more geodesics.  Lacking an
answer to the general question, however, we shall content ourselves with the
knowledge that for the metrics and null-geodesics we are interested in, namely
the ones corresponding to supersymmetric instantons, this
simultaneously-twistfree embedding is, as will be shown in the next section,
always possible.
\par
As we have seen so far, instanton solutions are associated only to the
existence of null geodesics of the $\sigma$-model metric. The relation between
instanton solutions and isometries allows us to define and compute an
instanton charge for the solution. Thus, we are lead to consider the cases in
which the $\sigma$-model metric is invariant under global transformations that
act on the adapted coordinates $\phi^{a-}$ as constant shifts.  The $a-$
components of the Killing vectors $k_{(a)}{}^{i}$ generating each of these
transformations will, thus, have the form $k_{(a)}{}^{b-} =
\delta_{a}{}^{b}$. The Noether currents associated to these invariances are,
in the adapted metric Eq.~(\ref{eq:adaptedmetric})

\begin{equation}
j_{(a)}{}^{\mu} = 
k_{(a)}{}^{i}\mathcal{G}_{ij}\partial^{\mu}\phi^{j}\, ,
\end{equation}

\noindent
and we can define the associated instanton charges $Q_{a}$ enclosed by a
3-dimensional hypersurface $\Sigma^{3}$ by the integrals

\begin{equation}
Q_{a} \equiv \int_{\Sigma^{(3)}}d\Sigma^{\mu}j_{(a)\, \mu}\, .   
\end{equation}

\noindent
These definitions can be rewritten as integrals over the 4-volume $V^{4}$
enclosed by $\Sigma^{(3)}$ $\partial V^{(4)}=\Sigma^{(3)}$:

\begin{equation}
Q_{a} = \int_{V^{(4)}}d^{4}x \sqrt{g}\,\nabla_{\mu} j_{(a)}{}^{\mu}\, .   
\end{equation}

These expressions do not vanish because the conservation of the Noether
currents is violated at the sources, which allows us to compute the charges
easily: since the Noether currents are conserved on shell, i.e.

\begin{equation}
\nabla_{\mu}j_{(a)}{}^{\mu} 
= 
\nabla_{\mu}(k_{(a)\, i}\partial^{\mu}\phi^{j})
= 
-\frac{k_{(a)}{}^{i}}{\sqrt{g}}\, \frac{\delta S_{\rm bulk}}{\delta \phi^{i}}\, .
\end{equation}

In presence of instanton sources the equations of motion of the scalars are

\begin{equation}
\frac{\delta S_{\rm bulk}}{\delta \phi^{i}}
+\frac{\delta S_{\rm inst}}{\delta \phi^{i}}=0\, ,  
\end{equation}

\noindent
and, using the instanton source in adapted coordinates
Eq.~(\ref{eq:adaptedinstantonsources})

\begin{equation}
\frac{\delta S_{\rm bulk}}{\delta \phi^{i}} = 
-\delta_{i}{}^{a-}\frac{f^{a}}{\sqrt{g}}\, ,  
\end{equation}

\noindent
so

\begin{equation}
\nabla_{\mu}j_{(a)}{}^{\mu} 
=
\frac{f^{a}}{\sqrt{g}}\, ,  
\end{equation}

\noindent
and

\begin{equation}
Q_{a} = \int_{V^{(4)}}d^{4}x f^{a}\, .   
\end{equation}

Thus, the source for $Q_{a}$ instantons of the species $a$ placed at
$x^{\mu}=x^{\mu}_{0}$ is just

\begin{equation}
f^{a}(x)= Q_{a}\delta^{(4)}(x-x_{0})\, .  
\end{equation}

We can try to evaluate naively the Euclidean action $S=S_{\rm bulk}+S_{\rm
  inst}$ for these instanton solutions. In principle one should add to this
action the Gibbons-Hawking boundary term found in Ref.~\cite{Gibbons:1976ue},
but its contribution is zero for flat Euclidean space \cite{Gibbons:1995vg}.
The bulk part of the action always vanishes on shell for gravity plus scalars
systems (irrespectively of their positive-definiteness) \cite{Kallosh:1992wa}
and we are left with

\begin{equation}
S= S_{\rm inst} = Q_{a}\phi^{a-}=  Q_{a}\phi^{a-}_{\infty}\, .  
\end{equation}

\noindent
($\phi^{a-}$, being constant, is also the value of $\phi^{a-}$ at infinity).

It has been argued in the literature (see,
e.g.~\cite{Bergshoeff:2008qq,Mohaupt:2009iq} that, in order to restore the
invariance of the total action $S=S_{\rm bulk}+S_{\rm inst}$ under the shifts
of $\phi^{a-}$, an appropriate boundary term ought to be introduced. In the
general case under consideration the only such term that can be introduced is

\begin{equation}
-\int d^{4}x\sqrt{g}\, \nabla_{\mu}(\phi^{a-}j_{(a)}{}^{\mu}) = 
-\int d^{3}\Sigma^{\mu}\phi^{a-}j_{(a)\, \mu}\, .  
\end{equation}

\noindent
Its contribution to the action, however, cancels identically that of the
source term: this was to be expected as the only way to recover the shift
invariance is to eliminate all explicit occurrences of $\phi^{a-}$ from the
result. 

On the other hand, in the cases of interest, the isometries that shift the
$\phi^{a-}$s also shift other coordinates, so the coordinates adapted to the
isometries do not coincide with the $\phi^{a-}$. Let $\chi^{a}(\phi)$ stand
for those adapted coordinates. The boundary term may then be

\begin{equation}
-\int d^{3}\Sigma^{\mu}\chi^{a}j_{(a)\, \mu}\, ,    
\end{equation}

\noindent
and then the Euclidean action would be given by the non-vanishing
shift-invariant result

\begin{equation}
S= Q_{a}(\phi^{a-}_{\infty}-\chi^{a}_{\infty})\, .  
\end{equation}

\subsection{Supersymmetric sourceless and sourceful instanton solutions}
\label{sec:SusyInst}

Let us now consider the case of interest for us: Wick-rotated $N=1,d=4$
supergravity coupled to chiral multiplets. The complex scalars $Z^{i}$ of the
Lorentzian theory consist of a real scalar and a pseudoscalar which, by
convention, we always take to be the real part of $Z^{i}$, something which is
always possible to achieve via field redefinitions. To Wick-rotate the
$Z^{i}$s we are going to use the rule of thumb/prescription that says that
pseudoscalars get an extra factor of $i$ in the Wick rotation so

\begin{equation}
Z^{i}\longrightarrow i Z^{i+}\, ,
\hspace{1cm}  
Z^{*i^{*}}\longrightarrow -i Z^{i-}\, ,
\end{equation}

\noindent
where $Z^{i+}$ and $Z^{i-}$ are two independent real scalars related to the
components of the complex scalars $Z^{i}$ by

\begin{equation}
Z^{i\pm} \equiv \Im {\rm m} (Z^{i})\pm  \Re {\rm e} (Z^{i})_{E}\, ,   
\end{equation}

\noindent
where the subscript $E$ indicates that we are dealing with the Wick-rotated
pseudoscalar. 

In many cases (in particular in the examples considered) this prescription
leads to Wick-rotated K\"ahler metrics $\mathcal{G}_{i^{+}j^{-}}$ which are
real\footnote{Actually \cite{Mohaupt:2009iq} these ``Wick-rotated K\"ahler
  metrics'' must be \textit{para-K\"ahler metrics}, which are real, split
  (signature $n,n$) metrics satisfying essentially the same properties as the
  K\"aherl metrics do but in terms of hyperbolic numbers which are generated
  over the reals by $1$ and $e$, where $e^{2}= +1$ and $e^{*}=-e$. Hyperbolic
  numbers take the general form $w=a + eb$ ($a,b \in \mathbb{R}$) and their
  conjugate is $w^{*} = a-eb$ so that $ww^{*}=a^{2}-b^{2}$.} and to the
Euclidean action

\begin{equation}
\label{eq:actionN1Euclidean}
 S  =  {\displaystyle\int} d^{4}x \sqrt{g}
\left[R 
+2\mathcal{G}_{i+j-}\partial_{\mu}Z^{i+}\partial^{\mu}Z^{j-} 
+2f^{a}(x)\mathcal{F}_{a}(Z^{+},Z^{-})\right]\, ,
\end{equation}

\noindent
where we have included a set of sources inspired in the general case with a
normalization adequate to our conventions for $N=1,d=4$ supergravity.  From
this action we get the equations of motion

\begin{eqnarray}
R_{\mu\nu} 
+2\mathcal{G}_{i+j-}\partial_{\mu}Z^{i+}\partial_{\nu}Z^{j-}  
& = & 
0\, ,
\\
& & \nonumber \\
\nabla^{2}Z^{i+}
+\Gamma_{j+k+}{}^{i+}\partial_{\mu}Z^{j+} \partial^{\mu}Z^{k+}
-f^{a}(x)\mathcal{G}^{i+j-}\partial_{j-}\mathcal{F}_{a}
& = & 
0\, ,
\\
& & \nonumber \\
\nabla^{2}Z^{i-} +\Gamma_{j-k-}{}^{i-}\partial_{\mu}Z^{j-} \partial^{\mu}Z^{k-}
-f^{a}(x)\mathcal{G}^{i-j+}\partial_{j+}\mathcal{F}_{a}
& = & 
0\, .
\end{eqnarray}

In order to have a flat spatial metric we must have 

\begin{equation}
\mathcal{G}_{i+j-}\partial_{\mu}Z^{i+}\partial_{\nu}Z^{j-}=0\, .
\end{equation}

\noindent
Inspired by the results of the general case, to solve this constraint we
assume that in the instanton solutions only the $Z^{i+}$s are nontrivial,
depending on up to $n$ functions $\sigma^{a}(x)$, while the $Z^{i-}$s are
constant\footnote{Interchanging everywhere indices $+$ and $-$ we go from
  instantons to anti-instantons.}:

\begin{equation}
\label{eq:ansatz}
\partial_{\mu}Z^{i+} = \partial_{\mu}\sigma^{a}\partial_{a}Z^{i+}\, ,
\hspace{1cm}  
\partial_{\mu}Z^{i-}=0\, .
\end{equation}

This Ansatz (called the ``extremal instanton Ansatz'' in
Ref.~\cite{Mohaupt:2009iq}) automatically solves the third equation of motion
if

\begin{equation}
\label{eq:Fcondition}
\partial_{j+}\mathcal{F}_{a}=0\, ,  
\end{equation}

\noindent
a result which will be shown to be consistent with the supersymmetry of the
source. After contraction with $\mathcal{G}_{l-i+}$, the second reduces to

\begin{equation}
\nabla^{2}\sigma^{a}\mathcal{G}_{l-i+}\partial_{a}Z^{i+} 
+\partial_{(a}(\mathcal{G}_{l-i+}\partial_{b)}Z^{i+}) 
\partial_{\mu}\sigma^{a}\partial_{\mu}\sigma^{b}
-f^{a}(x)\mathcal{G}^{i+j-}\partial_{j-}\mathcal{F}_{a}
=  
0\, ,  
\end{equation}

\noindent
which are solved by imposing, separately,

\begin{eqnarray}
\nabla^{2}\sigma^{a} 
& = & 
f^{a}(x)\, ,
\\
& & \nonumber \\  
\partial_{(a}(\mathcal{G}_{l-i+}\partial_{b)}Z^{i+}) 
& = & 
0\, ,
\end{eqnarray}

\noindent
and the consistency constraint

\begin{equation}
\label{eq:constraint}
\mathcal{G}_{i-j+}\partial_{a}Z^{i+} = \partial_{i-}\mathcal{F}_{a}\, .  
\end{equation}

\noindent
Now, given the K\"ahler origin of the metric,
$\mathcal{G}_{l-i+}=\partial_{i+}\partial_{l-}\mathcal{K}_{E}$, where
$\mathcal{K}_{E}$ is the Wick-rotated K\"ahler potential, and the last equation
reduces to 

\begin{equation}
\partial_{a}\partial_{b}\partial_{i-}\mathcal{K}_{E}=0\, ,
\end{equation}

\noindent
which can be integrated immediately

\begin{equation}
\partial_{i-}\mathcal{K}_{E} = c_{i} +d_{ia}\sigma^{a}\, ,  
\end{equation}

\noindent
for some integration constants $c_{i},d_{ia}$.  These are $n$ algebraic
equations involving the $\phi^{i+}(\sigma)$ and the constants $\phi^{j-}$ and
which, in principle, one should be able to solve for the $n$
$\phi^{i+}(\sigma)$. 

The constraint Eq.~(\ref{eq:constraint}) for the sources can also be solved
immediately:

\begin{equation}
\mathcal{F}_{a} = d_{ai}Z^{i-}\, .  
\end{equation}

At this point it is straightforward to go to adapted/Walker coordinates by
effecting the coordinate-transformation\footnote{ Observe that this coordinate
  transformation is in general not para-holomorphic but that it is always
  invertible.  }
\begin{equation}
  \label{eq:WalkerSUSY}
  \left.\begin{array}{lcl}
  \phi^{i-} &\equiv & Z^{i-} \\
  \phi^{+}_{i} &\equiv & \partial_{i-}\mathcal{K}_{E}
  \end{array}\right\} \;\longrightarrow\;
   \mathcal{G}_{i+j-}dZ^{i+}dZ^{j-}
   \ =\ d\phi^{i-}\ \left(\ d\phi_{i}^{+}\ +\ \mathsf{H}_{ij}\ d\phi^{j-}\right)\; ,
\end{equation}
whence all the information of the metric resides in 
\begin{equation}
  \label{eq:WalkerH}
  \mathsf{H}_{ij}\; \equiv\; \left. \partial_{i-}\partial_{j-}\mathcal{K}_{E}\right|_{Z^{\pm}=Z^{\pm}(\phi )}\; ,
\end{equation}
and generically depends on $\phi^{-}$ and $\phi^{+}$.

Let us then consider the issue of the unbroken supersymmetry of these
solutions: this is actually a worrisome point, one discussed at length in the
literature, as it concerns the Wick rotation of spinors.  The trouble is
easily recognised by seeing that in Lorentzian signature there exists in
4-dimensions a spinor with four real supercharges, in fact the one used to
build $N=1$ $d=4$ sugra, whereas in Euclidean signature the minimal spinor has
eight supercharges.  A consistent scheme for doing the Wick rotation was
developed in \cite{vanNieuwenhuizen:1996tv}, which does not solve the doubling
problem, but leaves the form of the supersymmetry transformations invariant,
which we hold to be a desireable property.  So accepting the doubling of
fermions, the gravitino, the chiralini and the supersymmetry parameter
$\psi_{\mu},\chi^{i},\epsilon$ will be rotated into
$\psi_{\mu}^{+},\chi^{i+},\epsilon^{+}$ whereas their complex conjugates
$\psi_{\mu}^{*},\chi^{i*},\epsilon^{*}$ will be rotated into {\em independent}
spinors $\psi_{\mu}^{-},\chi^{i-},\epsilon^{-}$. Then, the chiralini
supersymmetry rules give rise to the Killing spinor equations

\begin{equation}
\begin{array}{rcl}
\not\!\partial Z^{i+}\epsilon^{-} & = & 0\, ,
\\
& & \\
\not\!\partial Z^{i-}\epsilon^{+} & = & 0\, ,  
\\
\end{array}
\end{equation}

\noindent
which for the Ansatz Eq.~(\ref{eq:ansatz}), can be solved by setting

\begin{equation}
\label{eq:Killingspinorcondition}
\epsilon^{-}=0\, .  
\end{equation}

The gravitino supersymmetry transformation rule gives rise to another two
Killing spinor equations:

\begin{equation}
\begin{array}{rcl}
\left[\nabla_{\mu} 
+{\textstyle\frac{1}{4}}
(\partial_{\mu}Z^{i+}\partial_{i+}\mathcal{K}_{E}
-\partial_{\mu}Z^{i-}\partial_{i-}\mathcal{K}_{E}
)\right]\epsilon^{+} 
& = & 0\, ,
\\
& & \\ 
\left[\nabla_{\mu} 
-{\textstyle\frac{1}{4}}
(\partial_{\mu}Z^{i+}\partial_{i+}\mathcal{K}_{E}
-\partial_{\mu}Z^{i-}\partial_{i-}\mathcal{K}_{E}
)\right]\epsilon^{-}    
& = & 0\, .
\\
\end{array}
\end{equation}

\noindent
The second equation is solved by the condition
Eq.~(\ref{eq:Killingspinorcondition}) and the first can be rewritten in the
form

\begin{equation}
e^{-\mathcal{K}_{E}/4}\partial_{\mu}(e^{\mathcal{K}_{E}/4}\epsilon^{+})=0\, ,  
\end{equation}

\noindent
and is solved by 

\begin{equation}
\epsilon^{+}= e^{-\mathcal{K}_{E}/4}\epsilon^{+}_{0}\, ,  
\end{equation}

\noindent
for an arbitrary constant spinor $\epsilon^{+}_{0}$. 

Since we have only used the Ansatz Eq.~(\ref{eq:ansatz}) and not the particular
form of any solution, all the D-like instanton solutions preserve $1/2$ of the
supersymmetries.

Let us now consider the supersymmetry of the source. The supersymmetry
transformation rules of the scalars are

\begin{equation}
\delta_{\epsilon} Z^{i\pm} =
{\textstyle\frac{1}{4}} \bar{\chi}^{i\pm}\epsilon^{\pm}\, ,
\end{equation}

\noindent
and, thus, the supersymmetry variation of the source is

\begin{equation}
f^{a}(x)\{\partial_{i-}\mathcal{F}_{a}\delta_{\epsilon} Z^{i-}   
+
\partial_{i+}\mathcal{F}_{a}\delta_{\epsilon} Z^{i+}\}
=   
\tfrac{1}{4}
\{\partial_{i+}\mathcal{F}_{a}\bar{\chi}^{i+}\epsilon^{+}   
+
\partial_{i-}\mathcal{F}_{a}\bar{\chi}^{i-}\epsilon^{-}\}\, ,
\end{equation}

\noindent
and vanishes for either $\partial_{i+}\mathcal{F}_{a}=\epsilon^{-}=0$ (our
choice) or $\partial_{i-}\mathcal{F}_{a}=\epsilon^{+}=0$. We recover, then,
the condition Eq.~(\ref{eq:Fcondition}).

\subsubsection{Example 1:  $\mathrm{Sl}(2,\mathbb{R})/\mathrm{SO}(2)$}
\label{sec-instantonexamples1}

In this section we are going to consider the
$\mathrm{Sl}(2,\mathbb{R})/\mathrm{U}(1)$ $\sigma$-model which is ubiquitous
in supergravity theories\footnote{A D-like instanton solution in $N=1,d=4$
  string compactifications was first constructed in Ref.~\cite{Rey:1989xj}
  using the Kalb-Ramond 2-form instead of the dual pseudscalar field.}. Using
the standard coordinate $\tau \equiv \chi +ie^{-\phi}$ which takes values in
the upper half complex plane, the K\"ahler potential is $\mathcal{K}=
-\log\left(\Im{\rm m}\tau\right)$ and the kinetic term takes the form

\begin{equation}
2\mathcal{G}_{ij^{*}}\partial_{\mu}Z^{i} \partial^{\mu}Z^{*j^{*}}
=\tfrac{1}{2} 
\frac{\partial_{\mu}\tau \partial^{\mu}\tau^{*}}{(\Im{\rm m} \tau)^{2}}\, . 
\end{equation}

\noindent
$\mathrm{Sl}(2,\mathbb{R})$ acts on $\tau$ via fractional-linear transformations

\begin{equation}
\tau^{\prime}\ =\ \frac{a\tau +b}{c\tau +d}\, ,
\hspace{1cm}
ad-bc\, =\, 1\, ,  
\end{equation}

\noindent
which leave the target-space metric invariant. These transformations are
generated by 3 real Killing vectors $K_{T},K_{D},K_{K}$

\begin{equation}
K_{T} \ =\ \partial_{\chi}\, ,
\hspace{1cm}  
K_{D} \ =\ \partial_{\phi} \-\ \chi\partial_{\chi}\, ,
\hspace{1cm}  
K_{K} \ =\ \chi \partial_{\phi} 
\ -\ \tfrac{1}{2}(\chi^{2}-e^{-2\phi})\partial_{\chi}\, ,
\end{equation}

\noindent
which satisfy the algebra

\begin{equation}
[D,T] \ =\ T\, ,  
\hspace{1cm}  
[D,K] \ =\ -K\, ,  
\hspace{1cm}  
[T,K] \ =\ D\, .  
\end{equation}

\noindent
The corresponding components of the holomorphic Killing vectors
($K=k^{\tau}(\tau)\partial_{\tau}+\mathrm{c.c.}$) are

\begin{equation}
k_{T}{}^{\tau} = 1\, ,
\hspace{1cm}  
k_{D}{}^{\tau} = -\tau\, ,
\hspace{1cm}  
k_{K}{}^{\tau} = -\tfrac{1}{2}\tau^{2}\, ,
\end{equation}

\noindent
and the corresponding momentum maps\footnote{These are defined, for each
  isometry, by the two relations:
  \begin{equation}
  k_{\tau^{*}} =i\partial_{\tau^{*}}\mathcal{P}\, ,  
\hspace{1cm}
k{}^{\tau}\partial_{\tau} \mathcal{K} = i\mathcal{P} +\lambda\, .
  \end{equation}
}
are given by 

\begin{equation}
\mathcal{P}_{T} = -\frac{1}{2\Im{\rm m}\tau}\, ,
\hspace{1cm}  
\mathcal{P}_{D} = -\frac{\Re{\rm e} \tau}{2\Im{\rm m}\tau}\, ,
\hspace{1cm}  
\mathcal{P}_{K} = -\frac{|\tau|^{2}}{4\Im{\rm m}\tau}\, ,
\end{equation}  

\noindent
and

\begin{equation}
\lambda_{T} = 0\, ,
\hspace{1cm}  
\lambda_{D} = \frac{1}{2}\, ,
\hspace{1cm}  
\lambda_{K} = \tfrac{1}{2}\tau\, .
\end{equation}  

The complex coordinate $\tau$ is adapted to the isometry $T$, under which it
transforms by a real shift $\tau^{\prime} = \tau +b$; this clearly affects
only the real component $\chi$, which can be identified as the pseudoscalar
field. If we Wick-rotate $\tau$ according to the prescription we have given
$\chi\longrightarrow i\chi_{E}$ and

\begin{equation}
\tau \longrightarrow i \tau^{+} 
\equiv 
i (e^{-\phi} +\chi_{E})
\hspace{1cm}  
\tau^{*} \longrightarrow -i \tau^{-} 
\equiv 
-i (e^{-\phi} -\chi_{E})
\, ,
\end{equation}

\noindent
so\footnote{In the Wick rotation we go from our mostly-minus metric to a
  Euclidean mostly plus metric. This gives rise to a global sign in the scalar
  kinetic term as well as in the Ricci scalar.}

\begin{equation}
\tfrac{1}{2} 
\frac{\partial_{\mu}\tau \partial^{\mu}\tau^{*}}{(\Im{\rm m} \tau)^{2}}
\longrightarrow 
-\frac{2\partial_{\mu}\tau^{+} \partial^{\mu}\tau^{-}}{(\tau^{+} +\tau^{-})^{2}}
\, .  
\end{equation}

Including an instanton source associated to the coordinate $\tau^{-}$ the
Euclidean action takes the form

\begin{equation}
\label{eq:Tsourcefulaction}
 S_{T} \; =\;  {\displaystyle\int} d^{4}x \sqrt{g}
\left[R 
+\frac{2\partial_{\mu}\tau^{+} \partial^{\mu}\tau^{-}}{(\tau^{+} +\tau^{-})^{2}}
+2d_{T}f^{T}(x)\ \tau^{-}\ \right]\, ,
\end{equation}

\noindent
where $d_{T}$ is a constant that can be set to 1. 

The Noether current associated to the constant shifts $\chi_{E}\rightarrow
\chi_{E}-c$, $\tau^{\pm}\rightarrow \tau^{\pm}\mp c$ is 

\begin{equation}
j_{T}{}^{\mu} =
2\frac{\partial^{\mu}(\tau^{+}-\tau^{-})}{(\tau^{+}+\tau^{-})^{2}}\, ,  
\end{equation}

\noindent
and 

\begin{equation}
\partial_{\mu}\ (\sqrt{g}\, j_{T}{}^{\mu}) \; =\;
-\frac{\delta S_{\rm bulk}}{\delta \tau^{+}}
+\frac{\delta S_{\rm bulk}}{\delta \tau^{-}}\, ,  
\end{equation}

\noindent
which on-shell gives

\begin{equation}
  \partial_{\mu}(\sqrt{g}\, j_{T}{}^{\mu}) 
    \; =\; -\frac{\delta S_{\rm inst}}{\delta \tau^{-}} 
    \; =\;  2f^{T}(x)\, .  
\end{equation}

For the sake of simplicity we will take $f^{T}(x)=
\tfrac{1}{2}Q_{T}\delta^{(4)}(x)$ so that we will consider a single instanton source of
charge $Q_{T}$ placed at the origin.

Following the general procedure (``extremal instanton Ansatz''), the instanton
solutions are given by ($d_{T}=1$)

\begin{equation}
\begin{array}{rcl}
\tau^{-} 
& = & 
\mathrm{constant}\, ,
\\
& & \\ 
\partial_{\tau^{-}}\mathcal{K}_{E} 
& = & 
c_{T}+\sigma^{T}\, ,
\\
& & \\ 
\nabla^{2}\sigma^{T}
& = & 
\tfrac{1}{2}Q_{T} \delta^{(4)}(x)\, ,
\end{array}
\end{equation}

\noindent
where the para-K\"ahler potential is given by

\begin{equation}
\mathcal{K}_{E} = -\log{(\tau^{+}+\tau^{-})}\, .  
\end{equation}

\noindent
The additive constant $c_{T}$ can be absorbed into the harmonic function
$\sigma^{T}$ and we define

\begin{equation}
H_{T} \equiv -(c_{T}+\sigma^{T}) = 
-c_{T}+ \frac{1}{4\pi^{2}}\frac{Q_{T}/2}{r^{2}}\, .  
\end{equation}

\noindent
The second equation can be solved for $\tau^{+}$ as a function of the harmonic
function $H_{T}$ and the constants $c_{T},\tau^{-}$:

\begin{equation}
  \tau^{+} = -\tau^{-} +H_{T}^{-1}\, .
\end{equation}

\noindent
We can determine the constants $c_{T},\tau_{-}$ in terms of $\chi_{E\,
  \infty}$ and $\phi_{\infty}$ (the asymptotic values of $\chi_{E}$ and
$\phi$) and we find that the instanton solution can be written in the final
form

\begin{equation}
\begin{array}{rcl}
\tau^{-} 
& = & 
e^{-\phi_{\infty}}-\chi_{E\, \infty}\, ,
\\
& & \\ 
\tau^{+}
& = & 
-\tau^{-} +
2e^{-\phi_{\infty}}
\displaystyle{
\left(1 +\frac{1}{4\pi^{2}}\frac{e^{-\phi_{\infty}}Q_{T}}{r^{2}} 
\right)^{-1}}\, .
\\
\end{array}
\end{equation}

Let us now compute the Euclidean action of this instanton solution. According
to the general discussion, the action Eq.~(\ref{eq:Tsourcefulaction})
evaluated on the above instanton solution gives $\tau^{-}_{\infty}Q_{T}$ and
adding the total-derivative term

\begin{equation}
\int d^{4}x \partial_{\mu}[\sqrt{g}\, \tfrac{1}{2}(\tau^{+}-\tau^{-})j_{T}{}^{\mu}]  
\end{equation}

\noindent
needed to restore the shift-invariance, we get the standard result
$\frac{1}{2}(\tau^{+}_{\infty}-\tau^{-}_{\infty})Q_{T}
=e^{-\phi_{\infty}}Q_{T}$. \cite{Gibbons:1995vg}

Observe that, to obtain the instanton solution, we could have worked in the
adapted coordinates $\phi^{+},\phi^{-}$ defined by 

\begin{equation}
\tau^{+} \ \equiv\ (\phi^{+})^{-1}-\phi^{-}\;\; ,\;\;
\tau^{-} \ \equiv\ \phi^{-}\, ,
\hspace{.5cm}\longrightarrow\hspace{.5cm}
-\frac{d\tau^{+} d\tau^{-}}{(\tau^{+} +\tau^{-})^{2}}
 \ =\
d\phi^{+}d\phi^{-} 
+(\phi^{+})^{2}(d\phi^{-})^{2}\, .
\end{equation}

\noindent
Observe, however, that the reparametrization necessary to go to adapted
coordinates does not respect the para-Hermitean structure (it mixes $\tau^{+}$
and $\tau^{-}$).  According to the general discussion, it is clear that
$\phi^{+}$ is a geodesic coordinate and an instanton solution is provided by

\begin{equation}\label{eq:Sl2Sol}
\phi^{+} \; =\; \sigma^{T}\, ,
\hspace{1cm}
\phi^{-} \; =\; \mathrm{constant}\, .
\end{equation}

Adding a constant $c_{T}$ to the $\sigma^{T}$ we chose before, we recover
exactly the same solution. 

Now, however, we cannot easily recover the standard result for the Eclidean
action since, in order to restore shift invariance, the term that we should
add cancels automatically the contribution of the rest of the action because,
in these non-para-holomorphic coordinates, only $\phi^{-}$ transforms under
the shifts.

Let us now consider the instantons related to the isometry $D$. The complex
coordinate adapted to this isometry, $\xi$, is implicitly defined by
$k_{D}{}^{\tau}\partial_{\tau}=\partial_{\xi}$, so\footnote{The factor $i$
  has been chosen for consistency as will be explained.} $\tau=
ie^{-\xi}$ and the $\sigma$-model metric takes the form

\begin{equation}
\tfrac{1}{2} 
\frac{\partial_{\mu}\tau \partial^{\mu}\tau^{*}}{(\Im{\rm m} \tau)^{2}}
=
\tfrac{1}{2} 
\frac{\partial_{\mu}\xi \partial^{\mu}\xi^{*}}{\sin^{2}\left(\Im{\rm m}
  \xi\right)}\, .
\end{equation}

Under the $D$-transformations $\xi^{\prime} =\xi +c $ while
$\tau^{\prime}=e^{-c}\tau$.

In this case, it seems reasonable to take the imaginary part of $\xi$ to be
the pseudoscalar field, as otherwise the Wick rotation of the K\"ahler
potential would not be real. Furthermore, we find that, thanks to the $-i$
factor in the relation between $\tau$ and $\xi$

\begin{equation}
  \begin{array}{rcl}
\Re {\rm e} \tau & = & -e^{-\Re {\rm e} \xi}\sin{\Im {\rm m} \xi}\, ,  \\
& & \\
\Im {\rm m} \tau & = & e^{-\Re {\rm e} \xi}\cos{\Im {\rm m} \xi}\, ,  \\
\end{array}
\end{equation}

\noindent
so it will be consistent to take $\Re {\rm e} \tau$ and $\Im {\rm m} \xi$
to be pseudoscalars simultaneously.

We, then, define the complex coordinate $Z = i\xi$ and Wick-rotate it
using the standard prescription

\begin{equation}
Z\longrightarrow iZ^{+}\, ,
\hspace{1cm}
Z^{*}\longrightarrow -iZ^{-}\, ,  
\hspace{1cm}
Z^{\pm} = \Im {\rm m}Z \pm \Re {\rm e}Z_{E}\, ,
\end{equation}

\noindent
so we end up with 

\begin{equation}
-\tfrac{1}{2} 
\frac{\partial_{\mu}Z^{+} \partial^{\mu}Z^{-}}{\sinh^{2}(Z^{+}-Z^{-})/2}\, .  
\end{equation}

By performing the coordinate transformation

\begin{equation}
\label{eq:Z+Z-}
Z^{+}\; =\; 
  2\, \mathrm{arccoth}(\phi^{+})
  \ +\ 2\phi^{-}\, ,
\hspace{1cm}
Z^{-}\; =\; 2\phi^{-}\, ,  
\end{equation}

\noindent
we end up with the metric in the Walker form

\begin{equation}
2d\phi^{-}\ \left(\ d\phi^{+} \; -\; [(\phi^{+})^{2}-1]\ d\phi^{-}\ \right) \, .
\end{equation}

In these coordinates the instanton solution is again given by
Eq.~(\ref{eq:Sl2Sol}) and we just have to plug these expressions into
those of $Z^{+}$ and $Z^{-}$ in Eq.~(\ref{eq:Z+Z-}) to find $\Re{\rm e} Z_{E}$
and $\Im {\rm m}Z$.

Undoing the coordinate transformations, we find complicated expressions
for the original variables:

\begin{equation}
  \begin{array}{rcl}
   \chi_{E} & = & (\Re {\rm e}\tau)_{E} 
     \; =\; e^{-(Z^{+}+Z^{-})/2}\ \sinh (Z^{+}-Z^{-})/2\, ,\\
& & \\ 
   e^{-\phi} & = & \Im {\rm m}\tau 
     \; =\;  e^{-(Z^{+}+Z^{-})/2}\ \cosh (Z^{+}-Z^{-})/2\, ,\\
  \end{array}
\end{equation}

\subsubsection{Example 2:  Instantons on $\overline{\mathbb{CP}}^{n}$}
\label{sec-instantonexamples2}

This $\sigma$-model is defined by the K\"ahler potential

\begin{equation}
\mathcal{K}= -\log{(1-|Z|^{2})}\, ,
\hspace{1cm}
|Z|^{2} \equiv Z^{i}Z^{*i^{*}}\, ,  
\end{equation}

\noindent
which leads to the Fubini-Study metric

\begin{equation}
ds^{2}=2\mathcal{G}_{ij^{*}}dZ^{i}dZ^{*j^{*}}=
2\frac{dZ^{i}dZ^{*i^{*}}}{1-|Z|^{2}}
+2  \frac{Z^{i}Z^{*j^{*}}dZ^{i}dZ^{*j^{*}}}{(1-|Z|^{2})^{2}}\, .
\end{equation}

The standard prescripton for the Wick rotation 

\begin{equation}\label{eq:SlnWick}
Z^{i}\longrightarrow iZ^{i+}\, ,
\hspace{1cm}
Z^{*i^{*}}\longrightarrow -iZ^{i-}\, ,  
\hspace{1cm}
Z^{i\pm} = \Im {\rm m}Z^{i} \pm \Re {\rm e}Z^{i}_{E}\, ,
\end{equation}

\noindent
and the general discussion lead us to the Euclidean action

\begin{equation}
S  =  {\displaystyle\int} d^{4}x \sqrt{g}
\left[R 
+2\frac{\partial_{\mu}Z^{i+}\partial^{\mu}Z^{i-}}{1-Z^{i+}Z^{i-}}
+2\frac{Z^{i+}Z^{j-}\partial_{\mu}Z^{i+}
\partial^{\mu}Z^{j-}}{(1-Z^{i+}Z^{i-})^{2}}
+2d_{ia}f^{a}(x)Z^{i-}\right]\, .
\end{equation}

The Wick'ed scalar manifold is that of the symmetric space
$\mathrm{Sl}(n+1;\mathbb{R})/\left[\mathrm{SO}(1,1)\otimes\mathrm{Sl}(n;\mathbb{R})\right]$.
\par
Putting the instantons in the $Z^{+}$ directions we can go to the adapted
coordinate system by changing the coordinates as
\begin{equation}
  \label{eq:SlnCoord}
  Z^{i-}\; \equiv\; \phi^{i-}\hspace{.4cm},\hspace{.4cm}
  Z^{i+}\; \equiv\; \frac{\phi^{+}_{i}}{1\ +\ \phi^{+}\cdot\phi^{-}} \; ,
\end{equation}
where we have defined $\phi^{+}\cdot\phi^{-}\ =\ \phi^{+}_{i}\phi^{i-}$.
The metric in these Walker-coordinates is given by
\begin{equation}
  \label{eq:SlnWalker}
  \mathcal{G}_{i+j-}\ dZ^{i+}dZ^{j-}\; =\; d\phi^{+}\cdot d\phi^{-}
       \, -\,  \left( \phi^{+}\cdot d\phi^{-}\ \right)^{2}\; . 
\end{equation}
The above metric is adapted to the $n$ commuting and obvious Killing vectors
$k_{(i)}=\partial_{\phi^{i-}}$, and undoing the coordinate transformation in
Eq.~(\ref{eq:SlnCoord}), these Killing vectors are
\begin{equation}
  \label{eq:SlnWalKill}
  k_{(i)} \; =\; \partial_{\phi^{i-}} 
         \; =\; \partial_{i-}\, -\, Z^{i+}\ Z^{j+}\partial_{j+}\; .
\end{equation}
Observe that this Killing vector, will not lead to a holomorphic Killing
vector once we undo the Wick rotation (\ref{eq:SlnWick}). This situation can
be ameliorated by adding the $n$ commuting, para-conjugated Killing
vectors\footnote{ The fact that there are 2 sets of $n$ mutually commuting
  sets of isometries is not surprising as $\mathfrak{sl}(n+1;\mathbb{R})$
  admits the 3-grading
  $\mathfrak{L}_{-1}\oplus\mathfrak{L}_{0}\oplus\mathfrak{L}_{1}$, where
  $\mathfrak{L}_{0}\simeq
  \mathfrak{so}(1,1)\oplus\mathfrak{sl}(n;\mathbb{R})$.  }
\begin{eqnarray}
  \overline{k}_{(i)} & =& \partial_{i+}\; -\; Z^{i-}Z^{j-}\ \partial_{j-} \nonumber \\
    & =& \left( 1+\phi^{+}\cdot\phi^{-}\right)\ \partial_{\phi_{i}^{+}}
         \; +\; \phi^{i-}\ \left[ \phi^{+}\cdot\partial_{+}\ -\ \phi^{-}\cdot\partial_{-}\ \right] \; ,
\end{eqnarray}
where we introduced the abbreviation $\phi^{-}\cdot\partial_{-}\equiv
\phi^{j+}\partial_{\phi^{j-}}$ and similar for $\phi^{+}\cdot\partial_{+}$.


\section{Conclusions}
\label{sec:conclusions}

In this paper we have constructed effective actions for supersymmetric
conformally-flat domain walls and instantons of $N=1,d=4$ supergravity that
can be used as sources for the corresponding supersymmetric solutions, which
we have reviewed. In order to construct the domain-wall effective action we
have clarified the situation of the two 3-forms found in
Ref.~\cite{Hartong:2009az} that transform into the gravitino, showing that
there is indeed one deformation parameter associated to each of them.

In the domain-wall case we have seen how the consistent introduction of
domain-wall sources modifies the scalar potential via the local,
spacetime-dependent coupling constant to the superpotential. We have also seen
how the introduction of this local coupling constant in the supersymmetry
transformation rules leads to first order flow equations which imply the
second order equations of motion including the sources. Everything is
consistent with the existence of a fully supersymmetric and democratic action
for $N=1,d=4$ in which all the higher-rank potentials found in
Ref.~\cite{Hartong:2009az} are present and all the coupling constants are
local. The Killing spinor identities of such a democratic theory should imply
the mentioned relation between first- and second-order equations.

This work does not exahust the study of the effective actions of
supersymmetric objects of $N=1,d=4$ supergravity: those of strings have not
yet been constructed, although theyr may not be as interesting as in other
dimensions. Furthermore, in these theories there are supersymmetric domain
walls with $AdS_{3}$ worldvolumes whose effective actions may be different.

The construction of the effective actions of the 2- and 3-branes of $N=2,d=4$
supergravity (which are expected to exist) could prove very interesting since
we expect non-trivial intersections with strings and supersymmetric black
holes. However, it is necessary to find first the 3- and 4-form potentials of
these theories. Work in this direction is in progress.


\section*{Acknowledgments}

T.O.~would like to thank A.~Ach\'ucarro, J.~Hartong and J.~Urrestilla for very
useful conversations. This work has been supported in part by the Spanish
Ministry of Science and Education grants FPA2006-00783 and FPA2009-07692, the
Comunidad de Madrid grant HEPHACOS P-ESP-00346 and by the Spanish
Consolider-Ingenio 2010 program CPAN CSD2007-00042.  Further, TO wishes to
express his gratitude to M.M.~Fern\'andez for her permanent support.

\appendix


\end{document}